\documentclass[longauth]{aa}  

\usepackage{graphicx}
\usepackage{txfonts}
\usepackage{enumitem}   
\usepackage{multirow}   

\usepackage[colorlinks = true, linkcolor = blue, urlcolor  = blue, citecolor = blue]{hyperref}
\usepackage{euclid}
\usepackage{lipsum}
\usepackage{xcolor}

\definecolor{red1}{RGB}{180,1,25}
\definecolor{blue1}{RGB}{0,128,255}

\newcommand{\lob}{\lambda^{\rm ob}}

\newcommand{\zob}{z^{\rm ob}}

\defcitealias{Fumagalli21}{F21}
\defcitealias{Fumagalli-EP27}{EC:F24}

\begin{document} 

    \title{\Euclid: Survey mask effects on clustering of galaxy clusters\thanks{This paper is published on behalf of the Euclid Consortium.}}    
    \subtitle{The two-point correlation function monopole and its covariance}
    
\newcommand{\orcid}[1]{} 
\author{A.~Fumagalli\orcid{0009-0004-0300-2535}\thanks{\email{alessandra.fumagalli@inaf.it}}\inst{\ref{aff1}}
\and R.~Ingrao\orcid{0009-0009-6082-9818}\inst{\ref{aff2},\ref{aff3},\ref{aff1},\ref{aff4}}
\and M.~Costanzi\orcid{0000-0001-8158-1449}\inst{\ref{aff2},\ref{aff1},\ref{aff3}}
\and T.~Castro\orcid{0000-0002-6292-3228}\inst{\ref{aff5},\ref{aff1},\ref{aff3}}
\and S.~Borgani\orcid{0000-0001-6151-6439}\inst{\ref{aff2},\ref{aff3},\ref{aff1},\ref{aff4},\ref{aff6}}
\and A.~Saro\orcid{0000-0002-9288-862X}\inst{\ref{aff2},\ref{aff3},\ref{aff1},\ref{aff4},\ref{aff6}}
\and E.~Tsaprazi\orcid{0000-0001-5082-4380}\inst{\ref{aff7},\ref{aff8}}
\and F.~Marulli\orcid{0000-0002-8850-0303}\inst{\ref{aff9},\ref{aff10},\ref{aff11}}
\and M.~Aguena\orcid{0000-0001-5679-6747}\inst{\ref{aff1}}
\and L.~Baumont\orcid{0000-0002-1518-0150}\inst{\ref{aff2},\ref{aff1},\ref{aff3}}
\and N.~Aghanim\orcid{0000-0002-6688-8992}\inst{\ref{aff12}}
\and B.~Altieri\orcid{0000-0003-3936-0284}\inst{\ref{aff13}}
\and S.~Andreon\orcid{0000-0002-2041-8784}\inst{\ref{aff14}}
\and N.~Auricchio\orcid{0000-0003-4444-8651}\inst{\ref{aff10}}
\and C.~Baccigalupi\orcid{0000-0002-8211-1630}\inst{\ref{aff3},\ref{aff1},\ref{aff4},\ref{aff15}}
\and M.~Baldi\orcid{0000-0003-4145-1943}\inst{\ref{aff16},\ref{aff10},\ref{aff11}}
\and P.~Battaglia\orcid{0000-0002-7337-5909}\inst{\ref{aff10}}
\and A.~Biviano\orcid{0000-0002-0857-0732}\inst{\ref{aff1},\ref{aff3}}
\and E.~Branchini\orcid{0000-0002-0808-6908}\inst{\ref{aff17},\ref{aff18},\ref{aff14}}
\and M.~Brescia\orcid{0000-0001-9506-5680}\inst{\ref{aff19},\ref{aff20}}
\and J.~Brinchmann\orcid{0000-0003-4359-8797}\inst{\ref{aff21},\ref{aff22},\ref{aff23}}
\and S.~Camera\orcid{0000-0003-3399-3574}\inst{\ref{aff24},\ref{aff25},\ref{aff26}}
\and V.~Capobianco\orcid{0000-0002-3309-7692}\inst{\ref{aff26}}
\and C.~Carbone\orcid{0000-0003-0125-3563}\inst{\ref{aff27}}
\and M.~Castellano\orcid{0000-0001-9875-8263}\inst{\ref{aff28}}
\and G.~Castignani\orcid{0000-0001-6831-0687}\inst{\ref{aff10}}
\and S.~Cavuoti\orcid{0000-0002-3787-4196}\inst{\ref{aff20},\ref{aff29}}
\and A.~Cimatti\inst{\ref{aff30}}
\and C.~Colodro-Conde\inst{\ref{aff31}}
\and G.~Congedo\orcid{0000-0003-2508-0046}\inst{\ref{aff32}}
\and L.~Conversi\orcid{0000-0002-6710-8476}\inst{\ref{aff33},\ref{aff13}}
\and Y.~Copin\orcid{0000-0002-5317-7518}\inst{\ref{aff34}}
\and H.~M.~Courtois\orcid{0000-0003-0509-1776}\inst{\ref{aff35}}
\and H.~Degaudenzi\orcid{0000-0002-5887-6799}\inst{\ref{aff36}}
\and S.~de~la~Torre\inst{\ref{aff37}}
\and G.~De~Lucia\orcid{0000-0002-6220-9104}\inst{\ref{aff1}}
\and H.~Dole\orcid{0000-0002-9767-3839}\inst{\ref{aff12},\ref{aff38}}
\and F.~Dubath\orcid{0000-0002-6533-2810}\inst{\ref{aff36}}
\and X.~Dupac\inst{\ref{aff13}}
\and S.~Dusini\orcid{0000-0002-1128-0664}\inst{\ref{aff39}}
\and S.~Escoffier\orcid{0000-0002-2847-7498}\inst{\ref{aff40}}
\and M.~Farina\orcid{0000-0002-3089-7846}\inst{\ref{aff41}}
\and F.~Faustini\orcid{0000-0001-6274-5145}\inst{\ref{aff28}}
\and S.~Ferriol\inst{\ref{aff34}}
\and F.~Finelli\orcid{0000-0002-6694-3269}\inst{\ref{aff10},\ref{aff42}}
\and M.~Frailis\orcid{0000-0002-7400-2135}\inst{\ref{aff1}}
\and E.~Franceschi\orcid{0000-0002-0585-6591}\inst{\ref{aff10}}
\and M.~Fumana\orcid{0000-0001-6787-5950}\inst{\ref{aff27}}
\and L.~Gabarra\orcid{0000-0002-8486-8856}\inst{\ref{aff43}}
\and S.~Galeotta\orcid{0000-0002-3748-5115}\inst{\ref{aff1}}
\and K.~George\orcid{0000-0002-1734-8455}\inst{\ref{aff44}}
\and C.~Giocoli\orcid{0000-0002-9590-7961}\inst{\ref{aff10},\ref{aff11}}
\and A.~Grazian\orcid{0000-0002-5688-0663}\inst{\ref{aff45}}
\and F.~Grupp\inst{\ref{aff46},\ref{aff47}}
\and L.~Guzzo\orcid{0000-0001-8264-5192}\inst{\ref{aff48},\ref{aff14},\ref{aff49}}
\and S.~V.~H.~Haugan\orcid{0000-0001-9648-7260}\inst{\ref{aff50}}
\and S.~Hemmati\orcid{0000-0003-2226-5395}\inst{\ref{aff51}}
\and W.~Holmes\orcid{0009-0007-8554-4646}\inst{\ref{aff52}}
\and F.~Hormuth\inst{\ref{aff53}}
\and A.~Hornstrup\orcid{0000-0002-3363-0936}\inst{\ref{aff54},\ref{aff55}}
\and K.~Jahnke\orcid{0000-0003-3804-2137}\inst{\ref{aff56}}
\and M.~Jhabvala\inst{\ref{aff57}}
\and S.~Kermiche\orcid{0000-0002-0302-5735}\inst{\ref{aff40}}
\and A.~Kiessling\orcid{0000-0002-2590-1273}\inst{\ref{aff52}}
\and M.~Kilbinger\orcid{0000-0001-9513-7138}\inst{\ref{aff58}}
\and B.~Kubik\orcid{0009-0006-5823-4880}\inst{\ref{aff34}}
\and M.~K\"ummel\orcid{0000-0003-2791-2117}\inst{\ref{aff47}}
\and M.~Kunz\orcid{0000-0002-3052-7394}\inst{\ref{aff59}}
\and H.~Kurki-Suonio\orcid{0000-0002-4618-3063}\inst{\ref{aff60},\ref{aff61}}
\and A.~M.~C.~Le~Brun\orcid{0000-0002-0936-4594}\inst{\ref{aff8}}
\and S.~Ligori\orcid{0000-0003-4172-4606}\inst{\ref{aff26}}
\and P.~B.~Lilje\orcid{0000-0003-4324-7794}\inst{\ref{aff50}}
\and V.~Lindholm\orcid{0000-0003-2317-5471}\inst{\ref{aff60},\ref{aff61}}
\and I.~Lloro\orcid{0000-0001-5966-1434}\inst{\ref{aff62}}
\and M.~Magliocchetti\orcid{0000-0001-9158-4838}\inst{\ref{aff41}}
\and G.~Mainetti\orcid{0000-0003-2384-2377}\inst{\ref{aff63}}
\and S.~Marcin\inst{\ref{aff64}}
\and O.~Marggraf\orcid{0000-0001-7242-3852}\inst{\ref{aff65}}
\and M.~Martinelli\orcid{0000-0002-6943-7732}\inst{\ref{aff28},\ref{aff66}}
\and N.~Martinet\orcid{0000-0003-2786-7790}\inst{\ref{aff37}}
\and R.~J.~Massey\orcid{0000-0002-6085-3780}\inst{\ref{aff67}}
\and N.~Mauri\orcid{0000-0001-8196-1548}\inst{\ref{aff30},\ref{aff11}}
\and E.~Medinaceli\orcid{0000-0002-4040-7783}\inst{\ref{aff10}}
\and M.~Meneghetti\orcid{0000-0003-1225-7084}\inst{\ref{aff10},\ref{aff11}}
\and E.~Merlin\orcid{0000-0001-6870-8900}\inst{\ref{aff45}}
\and P.~Monaco\orcid{0000-0003-2083-7564}\inst{\ref{aff2},\ref{aff1},\ref{aff4},\ref{aff3}}
\and A.~Mora\orcid{0000-0002-1922-8529}\inst{\ref{aff68}}
\and M.~Moresco\orcid{0000-0002-7616-7136}\inst{\ref{aff9},\ref{aff10}}
\and L.~Moscardini\orcid{0000-0002-3473-6716}\inst{\ref{aff9},\ref{aff10},\ref{aff11}}
\and E.~Munari\orcid{0000-0002-1751-5946}\inst{\ref{aff1},\ref{aff3}}
\and C.~Neissner\orcid{0000-0001-8524-4968}\inst{\ref{aff69},\ref{aff70}}
\and S.-M.~Niemi\orcid{0009-0005-0247-0086}\inst{\ref{aff71}}
\and J.~W.~Nightingale\orcid{0000-0002-8987-7401}\inst{\ref{aff72}}
\and C.~Padilla\orcid{0000-0001-7951-0166}\inst{\ref{aff69}}
\and S.~Paltani\orcid{0000-0002-8108-9179}\inst{\ref{aff36}}
\and F.~Pasian\orcid{0000-0002-4869-3227}\inst{\ref{aff1}}
\and V.~Pettorino\orcid{0000-0002-4203-9320}\inst{\ref{aff71}}
\and A.~Pezzotta\orcid{0000-0003-0726-2268}\inst{\ref{aff14}}
\and S.~Pires\orcid{0000-0002-0249-2104}\inst{\ref{aff58}}
\and G.~Polenta\orcid{0000-0003-4067-9196}\inst{\ref{aff73}}
\and M.~Poncet\inst{\ref{aff74}}
\and L.~A.~Popa\inst{\ref{aff75}}
\and L.~Pozzetti\orcid{0000-0001-7085-0412}\inst{\ref{aff10}}
\and F.~Raison\orcid{0000-0002-7819-6918}\inst{\ref{aff46}}
\and A.~Renzi\orcid{0000-0001-9856-1970}\inst{\ref{aff76},\ref{aff39},\ref{aff10}}
\and J.~Rhodes\orcid{0000-0002-4485-8549}\inst{\ref{aff52}}
\and G.~Riccio\inst{\ref{aff20}}
\and I.~Risso\orcid{0000-0003-2525-7761}\inst{\ref{aff17},\ref{aff18},\ref{aff14}}
\and F.~Rizzo\orcid{0000-0002-9407-585X}\inst{\ref{aff1}}
\and E.~Romelli\orcid{0000-0003-3069-9222}\inst{\ref{aff1}}
\and R.~Saglia\orcid{0000-0003-0378-7032}\inst{\ref{aff47},\ref{aff46}}
\and A.~G.~S\'anchez\orcid{0000-0003-1198-831X}\inst{\ref{aff46}}
\and D.~Sapone\orcid{0000-0001-7089-4503}\inst{\ref{aff77}}
\and B.~Sartoris\orcid{0000-0003-1337-5269}\inst{\ref{aff78}}
\and M.~Schirmer\orcid{0000-0003-2568-9994}\inst{\ref{aff56}}
\and P.~Schneider\orcid{0000-0001-8561-2679}\inst{\ref{aff65}}
\and A.~Secroun\orcid{0000-0003-0505-3710}\inst{\ref{aff40}}
\and E.~Sihvola\orcid{0000-0003-1804-7715}\inst{\ref{aff60},\ref{aff61}}
\and P.~Simon\inst{\ref{aff65}}
\and C.~Sirignano\orcid{0000-0002-0995-7146}\inst{\ref{aff76},\ref{aff39}}
\and G.~Sirri\orcid{0000-0003-2626-2853}\inst{\ref{aff11}}
\and L.~Stanco\orcid{0000-0002-9706-5104}\inst{\ref{aff39}}
\and P.~Tallada-Cresp\'{i}\orcid{0000-0002-1336-8328}\inst{\ref{aff79},\ref{aff70}}
\and D.~Tavagnacco\orcid{0000-0001-7475-9894}\inst{\ref{aff1}}
\and A.~N.~Taylor\inst{\ref{aff32}}
\and I.~Tereno\orcid{0000-0002-4537-6218}\inst{\ref{aff80},\ref{aff81}}
\and R.~Toledo-Moreo\orcid{0000-0002-2997-4859}\inst{\ref{aff82},\ref{aff83}}
\and F.~Torradeflot\orcid{0000-0003-1160-1517}\inst{\ref{aff70},\ref{aff79}}
\and I.~Tutusaus\orcid{0000-0002-3199-0399}\inst{\ref{aff84},\ref{aff85},\ref{aff86}}
\and J.~Valiviita\orcid{0000-0001-6225-3693}\inst{\ref{aff60},\ref{aff61}}
\and T.~Vassallo\orcid{0000-0001-6512-6358}\inst{\ref{aff1},\ref{aff44}}
\and Y.~Wang\orcid{0000-0002-4749-2984}\inst{\ref{aff51}}
\and J.~Weller\orcid{0000-0002-8282-2010}\inst{\ref{aff47},\ref{aff46}}
\and G.~Zamorani\orcid{0000-0002-2318-301X}\inst{\ref{aff10}}
\and F.~M.~Zerbi\orcid{0000-0002-9996-973X}\inst{\ref{aff14}}
\and E.~Zucca\orcid{0000-0002-5845-8132}\inst{\ref{aff10}}
\and J.~Garc\'ia-Bellido\orcid{0000-0002-9370-8360}\inst{\ref{aff87}}
\and C.~M.~Gutierrez\orcid{0000-0001-7854-783X}\inst{\ref{aff31},\ref{aff88}}
\and M.~Sereno\orcid{0000-0003-0302-0325}\inst{\ref{aff10},\ref{aff11}}
\and K.~Tanidis\orcid{0000-0001-9843-5130}\inst{\ref{aff89}}
\and M.~Viel\orcid{0000-0002-2642-5707}\inst{\ref{aff3},\ref{aff1},\ref{aff15},\ref{aff4},\ref{aff6}}}
										   
\institute{INAF-Osservatorio Astronomico di Trieste, Via G. B. Tiepolo 11, 34143 Trieste, Italy\label{aff1}
\and
Dipartimento di Fisica - Sezione di Astronomia, Universit\`a di Trieste, Via Tiepolo 11, 34131 Trieste, Italy\label{aff2}
\and
IFPU, Institute for Fundamental Physics of the Universe, via Beirut 2, 34151 Trieste, Italy\label{aff3}
\and
INFN, Sezione di Trieste, Via Valerio 2, 34127 Trieste TS, Italy\label{aff4}
\and
Department of Mathematical Physics, Institute of Physics, University of S\~ao Paulo, R. do Mat\~ao 1371, 05508-090, S\~ao Paulo, SP, Brazil\label{aff5}
\and
ICSC - Centro Nazionale di Ricerca in High Performance Computing, Big Data e Quantum Computing, Via Magnanelli 2, Bologna, Italy\label{aff6}
\and
Astrophysics Group, Blackett Laboratory, Imperial College London, London SW7 2AZ, UK\label{aff7}
\and
Laboratoire d'etude de l'Univers et des phenomenes eXtremes, Observatoire de Paris, Universit\'e PSL, Sorbonne Universit\'e, CNRS, 92190 Meudon, France\label{aff8}
\and
Dipartimento di Fisica e Astronomia "Augusto Righi" - Alma Mater Studiorum Universit\`a di Bologna, via Piero Gobetti 93/2, 40129 Bologna, Italy\label{aff9}
\and
INAF-Osservatorio di Astrofisica e Scienza dello Spazio di Bologna, Via Piero Gobetti 93/3, 40129 Bologna, Italy\label{aff10}
\and
INFN-Sezione di Bologna, Viale Berti Pichat 6/2, 40127 Bologna, Italy\label{aff11}
\and
Universit\'e Paris-Saclay, CNRS, Institut d'astrophysique spatiale, 91405, Orsay, France\label{aff12}
\and
ESAC/ESA, Camino Bajo del Castillo, s/n., Urb. Villafranca del Castillo, 28692 Villanueva de la Ca\~nada, Madrid, Spain\label{aff13}
\and
INAF-Osservatorio Astronomico di Brera, Via Brera 28, 20122 Milano, Italy\label{aff14}
\and
SISSA, International School for Advanced Studies, Via Bonomea 265, 34136 Trieste TS, Italy\label{aff15}
\and
Dipartimento di Fisica e Astronomia, Universit\`a di Bologna, Via Gobetti 93/2, 40129 Bologna, Italy\label{aff16}
\and
Dipartimento di Fisica, Universit\`a di Genova, Via Dodecaneso 33, 16146, Genova, Italy\label{aff17}
\and
INFN-Sezione di Genova, Via Dodecaneso 33, 16146, Genova, Italy\label{aff18}
\and
Department of Physics "E. Pancini", University Federico II, Via Cinthia 6, 80126, Napoli, Italy\label{aff19}
\and
INAF-Osservatorio Astronomico di Capodimonte, Via Moiariello 16, 80131 Napoli, Italy\label{aff20}
\and
European Southern Observatory, Karl-Schwarzschild-Str.~2, 85748 Garching, Germany\label{aff21}
\and
Instituto de Astrof\'isica e Ci\^encias do Espa\c{c}o, Universidade do Porto, CAUP, Rua das Estrelas, PT4150-762 Porto, Portugal\label{aff22}
\and
Faculdade de Ci\^encias da Universidade do Porto, Rua do Campo de Alegre, 4150-007 Porto, Portugal\label{aff23}
\and
Dipartimento di Fisica, Universit\`a degli Studi di Torino, Via P. Giuria 1, 10125 Torino, Italy\label{aff24}
\and
INFN-Sezione di Torino, Via P. Giuria 1, 10125 Torino, Italy\label{aff25}
\and
INAF-Osservatorio Astrofisico di Torino, Via Osservatorio 20, 10025 Pino Torinese (TO), Italy\label{aff26}
\and
INAF-IASF Milano, Via Alfonso Corti 12, 20133 Milano, Italy\label{aff27}
\and
INAF-Osservatorio Astronomico di Roma, Via Frascati 33, 00078 Monteporzio Catone, Italy\label{aff28}
\and
INFN -- Sezione di Napoli, Via Cinthia 6, 80126, Napoli, Italy\label{aff29}
\and
Dipartimento di Fisica e Astronomia "Augusto Righi" - Alma Mater Studiorum Universit\`a di Bologna, Viale Berti Pichat 6/2, 40127 Bologna, Italy\label{aff30}
\and
Instituto de Astrof\'{\i}sica de Canarias, E-38205 La Laguna, Tenerife, Spain\label{aff31}
\and
Institute for Astronomy, University of Edinburgh, Royal Observatory, Blackford Hill, Edinburgh EH9 3HJ, UK\label{aff32}
\and
European Space Agency/ESRIN, Largo Galileo Galilei 1, 00044 Frascati, Roma, Italy\label{aff33}
\and
Universit\'e Claude Bernard Lyon 1, CNRS/IN2P3, IP2I Lyon, UMR 5822, Villeurbanne, F-69100, France\label{aff34}
\and
UCB Lyon 1, CNRS/IN2P3, IUF, IP2I Lyon, 4 rue Enrico Fermi, 69622 Villeurbanne, France\label{aff35}
\and
Department of Astronomy, University of Geneva, ch. d'Ecogia 16, 1290 Versoix, Switzerland\label{aff36}
\and
Aix-Marseille Universit\'e, CNRS, CNES, LAM, Marseille, France\label{aff37}
\and
Institut universitaire de France (IUF), 1 rue Descartes, 75231 PARIS CEDEX 05, France\label{aff38}
\and
INFN-Padova, Via Marzolo 8, 35131 Padova, Italy\label{aff39}
\and
Aix-Marseille Universit\'e, CNRS/IN2P3, CPPM, Marseille, France\label{aff40}
\and
INAF-Istituto di Astrofisica e Planetologia Spaziali, via del Fosso del Cavaliere, 100, 00100 Roma, Italy\label{aff41}
\and
INFN-Bologna, Via Irnerio 46, 40126 Bologna, Italy\label{aff42}
\and
Department of Physics, University of Oxford, Keble Road, Oxford OX1 3RH, UK\label{aff43}
\and
University Observatory, LMU Faculty of Physics, Scheinerstr.~1, 81679 Munich, Germany\label{aff44}
\and
INAF-Osservatorio Astronomico di Padova, Via dell'Osservatorio 5, 35122 Padova, Italy\label{aff45}
\and
Max Planck Institute for Extraterrestrial Physics, Giessenbachstr. 1, 85748 Garching, Germany\label{aff46}
\and
Universit\"ats-Sternwarte M\"unchen, Fakult\"at f\"ur Physik, Ludwig-Maximilians-Universit\"at M\"unchen, Scheinerstr.~1, 81679 M\"unchen, Germany\label{aff47}
\and
Dipartimento di Fisica "Aldo Pontremoli", Universit\`a degli Studi di Milano, Via Celoria 16, 20133 Milano, Italy\label{aff48}
\and
INFN-Sezione di Milano, Via Celoria 16, 20133 Milano, Italy\label{aff49}
\and
Institute of Theoretical Astrophysics, University of Oslo, P.O. Box 1029 Blindern, 0315 Oslo, Norway\label{aff50}
\and
Caltech/IPAC, 1200 E. California Blvd., Pasadena, CA 91125, USA\label{aff51}
\and
Jet Propulsion Laboratory, California Institute of Technology, 4800 Oak Grove Drive, Pasadena, CA, 91109, USA\label{aff52}
\and
Felix Hormuth Engineering, Goethestr. 17, 69181 Leimen, Germany\label{aff53}
\and
Technical University of Denmark, Elektrovej 327, 2800 Kgs. Lyngby, Denmark\label{aff54}
\and
Cosmic Dawn Center (DAWN), Denmark\label{aff55}
\and
Max-Planck-Institut f\"ur Astronomie, K\"onigstuhl 17, 69117 Heidelberg, Germany\label{aff56}
\and
NASA Goddard Space Flight Center, Greenbelt, MD 20771, USA\label{aff57}
\and
Universit\'e Paris-Saclay, Universit\'e Paris Cit\'e, CEA, CNRS, AIM, 91191, Gif-sur-Yvette, France\label{aff58}
\and
Universit\'e de Gen\`eve, D\'epartement de Physique Th\'eorique and Centre for Astroparticle Physics, 24 quai Ernest-Ansermet, CH-1211 Gen\`eve 4, Switzerland\label{aff59}
\and
Department of Physics, P.O. Box 64, University of Helsinki, 00014 Helsinki, Finland\label{aff60}
\and
Helsinki Institute of Physics, Gustaf H{\"a}llstr{\"o}min katu 2, University of Helsinki, 00014 Helsinki, Finland\label{aff61}
\and
SKAO, Jodrell Bank, Lower Withington, Macclesfield SK11 9FT, UK\label{aff62}
\and
Centre de Calcul de l'IN2P3/CNRS, 21 avenue Pierre de Coubertin 69627 Villeurbanne Cedex, France\label{aff63}
\and
University of Applied Sciences and Arts of Northwestern Switzerland, School of Computer Science, 5210 Windisch, Switzerland\label{aff64}
\and
Universit\"at Bonn, Argelander-Institut f\"ur Astronomie, Auf dem H\"ugel 71, 53121 Bonn, Germany\label{aff65}
\and
INFN-Sezione di Roma, Piazzale Aldo Moro, 2 - c/o Dipartimento di Fisica, Edificio G. Marconi, 00185 Roma, Italy\label{aff66}
\and
Department of Physics, Institute for Computational Cosmology, Durham University, South Road, Durham, DH1 3LE, UK\label{aff67}
\and
Telespazio UK S.L. for European Space Agency (ESA), Camino bajo del Castillo, s/n, Urbanizacion Villafranca del Castillo, Villanueva de la Ca\~nada, 28692 Madrid, Spain\label{aff68}
\and
Institut de F\'{i}sica d'Altes Energies (IFAE), The Barcelona Institute of Science and Technology, Campus UAB, 08193 Bellaterra (Barcelona), Spain\label{aff69}
\and
Port d'Informaci\'{o} Cient\'{i}fica, Campus UAB, C. Albareda s/n, 08193 Bellaterra (Barcelona), Spain\label{aff70}
\and
European Space Agency/ESTEC, Keplerlaan 1, 2201 AZ Noordwijk, The Netherlands\label{aff71}
\and
School of Mathematics, Statistics and Physics, Newcastle University, Herschel Building, Newcastle-upon-Tyne, NE1 7RU, UK\label{aff72}
\and
Space Science Data Center, Italian Space Agency, via del Politecnico snc, 00133 Roma, Italy\label{aff73}
\and
Centre National d'Etudes Spatiales -- Centre spatial de Toulouse, 18 avenue Edouard Belin, 31401 Toulouse Cedex 9, France\label{aff74}
\and
Institute of Space Science, Str. Atomistilor, nr. 409 M\u{a}gurele, Ilfov, 077125, Romania\label{aff75}
\and
Dipartimento di Fisica e Astronomia "G. Galilei", Universit\`a di Padova, Via Marzolo 8, 35131 Padova, Italy\label{aff76}
\and
Departamento de F\'isica, FCFM, Universidad de Chile, Blanco Encalada 2008, Santiago, Chile\label{aff77}
\and
Universit\'e C\^{o}te d'Azur, Observatoire de la C\^{o}te d'Azur, CNRS, Laboratoire Lagrange, Bd de l'Observatoire, CS 34229, 06304 Nice cedex 4, France\label{aff78}
\and
Centro de Investigaciones Energ\'eticas, Medioambientales y Tecnol\'ogicas (CIEMAT), Avenida Complutense 40, 28040 Madrid, Spain\label{aff79}
\and
Departamento de F\'isica, Faculdade de Ci\^encias, Universidade de Lisboa, Edif\'icio C8, Campo Grande, PT1749-016 Lisboa, Portugal\label{aff80}
\and
Instituto de Astrof\'isica e Ci\^encias do Espa\c{c}o, Faculdade de Ci\^encias, Universidade de Lisboa, Campo Grande, 1749-016 Lisboa, Portugal\label{aff81}
\and
Universidad Polit\'ecnica de Cartagena, Departamento de Electr\'onica y Tecnolog\'ia de Computadoras,  Plaza del Hospital 1, 30202 Cartagena, Spain\label{aff82}
\and
European University of Technology EUt+, European Union\label{aff83}
\and
Institute of Space Sciences (ICE, CSIC), Campus UAB, Carrer de Can Magrans, s/n, 08193 Barcelona, Spain\label{aff84}
\and
Institut d'Estudis Espacials de Catalunya (IEEC),  Edifici RDIT, Campus UPC, 08860 Castelldefels, Barcelona, Spain\label{aff85}
\and
Institut de Recherche en Astrophysique et Plan\'etologie (IRAP), Universit\'e de Toulouse, CNRS, UPS, CNES, 14 Av. Edouard Belin, 31400 Toulouse, France\label{aff86}
\and
Instituto de F\'isica Te\'orica UAM-CSIC, Campus de Cantoblanco, 28049 Madrid, Spain\label{aff87}
\and
Universidad de La Laguna, Dpto. Astrof\'\i sica, E-38206 La Laguna, Tenerife, Spain\label{aff88}
\and
Center for Astrophysics and Cosmology, University of Nova Gorica, Nova Gorica, Slovenia\label{aff89}}    

   \date{Received September 30, 20XX}

 
  \abstract
    {Wide-area cluster catalogues from Stage-IV surveys will substantially improve the precision of cluster clustering measurements, making the control of observational systematic effects essential for cosmological inference. In preparation for the first \Euclid data release (DR1), we assess the impact of the survey mask and non-homogeneous sky coverage on the two-point correlation function of galaxy clusters and its covariance. Using DR1-like lightcone mocks, we confirm that the irregular footprint does not introduce appreciable changes in the two-point correlation function when the same mask is applied to the random catalogue. We also find that spatial inhomogeneities, implemented as variations of the cluster selection function in different patches of the sky, do not significantly bias the clustering amplitude, up to levels of variation well above those expected for DR1. When the degree of inhomogeneity is increased to the regime where the effect becomes significant, we show that its amplitude is related to the maximum variation of the selection function among different regions, rather than to the number of such regions. We further find that irregular mask geometry and non-uniform sky coverage both increase the clustering covariance, introducing small, nearly scale-independent offsets that can be captured by nuisance parameters in the covariance model. Our results consolidate the treatment of mask systematic errors for DR1 and inform the methodological refinements required for future \Euclid cluster cosmology.}
    
   \keywords{galaxies: clusters: general -- large-scale structure of Universe -- cosmological parameters -- methods: statistical}

    \titlerunning{\Euclid: Survey mask and homogeneous regions on 2PCF}
    \authorrunning{A. Fumagalli et al.}

   \maketitle

\nolinenumbers
\section{Introduction}
The ongoing generation of wide-field cosmological surveys -- characterised by deeper observations, extensive sky coverage, and high-precision measurements -- will significantly advance our understanding of the large-scale structure of the Universe. 
Among cosmological probes, galaxy clusters serve as powerful tracers of the highest peaks in the matter density field \citep{Kravtsov:2012zs}. Because of this connection to the underlying matter distribution and evolution, they provide valuable cosmological information that Stage IV surveys will enable us to access with improved precision.
The ESA mission \Euclid \citep{Laureijs11, Scaramella-EP1, EuclidSkyOverview}, with its extensive optical and near-infrared coverage and the space-based imaging quality, will bring cluster cosmology into a regime where statistical uncertainties are greatly reduced and systematic uncertainties become a dominant limitation, underscoring the need for a robust and well tested framework. The first \Euclid data release (DR1), covering approximately 1900 $\deg^2$ of the $14\,000 \, \deg^2$ planned for the final release of the mission, will provide new cosmological measurements while also serving as a key testbed for evaluating and refining the cluster cosmology framework under realistic observational conditions.

While cluster number counts have traditionally formed the backbone of cluster cosmology \citep[e.g.,][]{Borgani:2001ir,DES:2020cbm,SPT:2024qbr,Ghirardini:2024yni,Lesci:2025bqt}, they only partially exploit the information available in these samples. The spatial distribution of clusters encodes additional constraints, linked to the matter power spectrum, halo bias, and large-scale structure features like baryon acoustic oscillations (BAO), which can improve parameter inference and help resolve degeneracies associated with mass calibration \citep{Borgani:1998sfa,Mana:2013qba,Sereno:2014eea,Fumagalli:2023yym,Romanello:2024kbo}. To fully exploit this information, however, the analysis of cluster clustering must be supported by a solid methodological framework tailored to their specific properties.

In a series of preparatory studies we addressed the role of various systematic uncertainties in shaping the two-point correlation function (2PCF) of clusters and its covariance. \citet{Fumagalli-EP27} investigated theoretical systematic uncertainties in the cluster 2PCF, such as those arising from shot-noise and non-Gaussian contributions, and modelled the associated covariance matrix. Building on this, \citet{Fumagalli25} examined the impact of photometric redshift (photo-$z$) errors and redshift-space distortions (RSDs) on the 2PCF monopole, highlighting the potential biases introduced when analyses rely on simplified assumptions. More recently, \citet{EP-Tsaprazi} compared different models to describe the 2PCF multipoles in photometric surveys, proving the sufficiency of the dispersion model for a DR1-like survey.

A central ingredient that has not yet been fully addressed is the impact of the survey mask. Realistic survey footprints contain complex shapes, holes, instrumental artefacts, and heterogeneous data quality regions that directly shape the observed cluster distribution. 
Survey mask effects are well established in galaxy clustering analyses \citep[e.g.,][]{SDSS:2009ocz,Sato:2013hea,Colavincenzo:2017yhx,Beauchamps:2021fhb}. In Fourier and harmonic space, the mask induces a non-trivial mode mixing that must be accounted for through explicit window function convolutions when modelling the power spectrum. In configuration space, the impact on the 2PCF is typically milder, provided that the angular selection function is accurately captured by the random catalogue. For cluster samples, however, the situation has not been examined with the same level of detail. Although the general strategy of using random catalogues that reproduce the survey properties applies equally to galaxy and cluster clustering, the lower number density of clusters leads to different statistical properties. Existing analyses \citep[e.g.,][]{DES:2020uce,Fumagalli:2023yym,Romanello:2024kbo} typically account for the mask only by applying it to the random catalogue, without quantifying how irregular geometry or spatially varying observational conditions propagate into cluster 2PCF measurements or their covariance. Because clusters are sparse and highly-biased tracers and their clustering signal is measured on large scales, distortions introduced by the mask can propagate non-trivially into pair counts and correlation function estimates. Properly accounting for these effects is therefore essential for ensuring that the clustering measurements expected from upcoming cluster catalogues can be interpreted robustly in a cosmological context. 

In this study, we investigate the impact of irregular mask geometry and non-uniform sky coverage on cluster clustering measurements in a DR1-like context. We test whether two simplifying assumptions in the clustering modelling -- a circular footprint and an averaged selection function across the whole footprint -- provide an adequate description of the 2PCF and its covariance without introducing significant biases.
In the present study, we focus on the impact of these effects on the monopole of the 2PCF and its covariance, leaving the investigation of higher order multipoles to future work. Beyond the main analysis presented here, we include in Appendix\,\ref{app:count_cov} an investigation of the sensitivity of the number count covariance to the survey mask, motivated by its dependence on the clustering signal.

This paper is structured as follows. In Sect.~\ref{sec:theory} we present the theoretical framework used to model the 2PCF and its covariance. In Sect.~\ref{sec:data} we describe the simulations used in this work (Sect.~\ref{sec:data:mocks}), along with the survey masks applied to the mocks (Sect.~\ref{sec:data:mask}) and the homogeneous regions considered in this work (Sect.~\ref{sec:data:regions}). In Sect.~\ref{sec:results} we present our results: the comparison of different survey masks (Sect.~\ref{sec:results:mask}), the comparison of different homogeneous regions (Sect.~\ref{sec:results:regions} and \ref{sec:results:regions:robustness}), and the fit of the covariance matrix model (Sect.\,\ref{sec:results:cov_fit}). In Sect.~\ref{sec:conclusions} we discuss our conclusions.

\section{Theoretical background} \label{sec:theory}
In this section we describe the modelling of the 2PCF monopole and its covariance. Since we consider a single richness bin and two independent redshift intervals (Sect.~\ref{sec:data:measure}), we restrict the analysis to the auto-correlation between radial separation bins.\footnote{Because only one richness bin is used, richness-bin indices are omitted in the equations.} 
We model the 2PCF in the $a$-th observed redshift bin of width $\Delta \zob_a$ and $i$-th radial bin of width $\Delta r_i$ as
\begin{equation} \label{eq:2pcf_obs}
    \xi(\Delta \zob_a, \Delta r_i) = \int \frac{\diff k\,k^2}{2\pi^2} \,P_{\rm h}(k, \Delta \zob_a)\, \, W_i(k)\,,
\end{equation}
where $W_i(k)$ is the window function of the spherical shells, given by
\begin{equation} \label{eq:shell_window}
     W_i(k) = \int \frac{\diff^3r}{V_i} j_0(kr) = \frac{r_{i,+}^3 W_{\rm th}(k r_{i,+}) - r_{i,-}^3 W_{\rm th}(k r_{i,-})}{r_{i,+}^3 - r_{i,-}^3}\,,
\end{equation}
where $V_i$ is the volume of the $i$-th spherical shell, delimited by $r_{i,-}$ and $r_{i,+}$, and $W_{\rm th}(kr)$ is the top-hat window function.

The term $P_{\rm h}(k, \Delta \zob_a)$ is the halo power spectrum integrated over the $a$-th redshift bin, given by
\begin{equation} \label{eq:pk_a_i}
    \begin{split}
	  P_{\rm h}(k, \Delta \zob_a) = \frac{1}{N(\Delta \zob_a)} \int_0^\infty \diff z \, \Omega_{\rm mask}\,&\frac{\diff V}{\diff \Omega\,\diff z}(z | \Delta \zob_a) \\
     & \times n^2(z) \, b^2_{\rm eff}(z) \,P_{\rm m}(k,z)\,,    
    \end{split}
\end{equation}
where $P_{\rm m}(k,z)$ is the linear matter power spectrum, $N(\Delta \zob_a)$ is the number of clusters in the observed redshift bin, $n(z)$ and $b_{\rm eff}(z)$ are the number density and the effective bias of halos, $\Omega_{\rm mask}$ is the survey mask area in steradians -- assuming homogeneous data properties over the footprint -- and the observed comoving volume element per unit redshift and solid angle is given by
\begin{equation} \label{eq:obs_vol}
\frac{\diff V}{\diff z \, \diff \Omega}(z|\Delta \zob_a) =  \int_{\Delta \zob_a} \diff \zob \frac{\diff V}{\diff z \, \diff \Omega}(z)\,{\cal P}(\zob\,|\,z)\,,
\end{equation}
where the term ${\cal P}(\zob \,|\, z)$ is the probability distribution that assigns an observed redshift $\zob$ to a cluster, given the true redshift. This term accounts for the effect of photo-$z$ errors that we model as a Gaussian distribution centred on $z$, with scatter
\begin{equation} \label{eq:photoz}
    \sigma_z(z) = \sigma_{z}^{\rm fit}\,(1+z)\,,
\end{equation}
where $\sigma_z^{\rm fit}$ is calibrated on the Flagship 2 simulation \citep[][]{EuclidSkyFlagship}, post-processed to reproduce the photometric properties of the Euclid Trial Release~1 (TR1). The photometry is replicated at the flux level following \citet[][their equation 5]{EP-Humphrey}, and photometric redshifts are assigned through nearest-neighbour matching in flux space with the TR1 catalogue (Euclid Collaboration: Murray et al. in prep., Euclid Collaboration: Aguena \& Baumont in prep.).
The fitting procedure follows the one adopted in the official \Euclid richness-assignment algorithm foreseen for the DR1 cosmological analysis, \texttt{RICH-CL-ZP} (Euclid Collaboration: Benoist et al. in prep.). For each galaxy, the photo-$z$ uncertainty is estimated from the width of the central 70 per cent interval of the redshift probability distribution (PDZ),
\[
    \sigma_0^{\rm PDZ} =
    \frac{z_{70,\max}-z_{70,\min}}{2(1+z_{\rm phot})}.
\]
The calibration first applies the survey magnitude limit corresponding to the nominal \Euclid NISP \HE-band depth (\HE$ \simeq 24.4$ AB; see, e.g., \citealt{Scaramella-EP1,EuclidSkyOverview}), and then imposes an $\HE < \HE^\star(z)+\Delta m$ selection, capped at the same limiting magnitude. Here, \HE is the galaxy apparent magnitude, $\HE^\star(z)$ is the characteristic magnitude at redshift $z$, and $\Delta m$ sets the allowed magnitude range below $\HE^\star(z)$. The median $\sigma_0^{\rm PDZ}$ is then measured in redshift bins and fitted with a third-order polynomial. The cluster photo-$z$ is then obtained by averaging the photometric redshifts of the member galaxies, weighted by their membership probabilities. In this work we use a redshift-only refit of this relation,
\begin{equation} \label{eq:sigma_z}
    \sigma_{z}^{\rm fit} = a_0 + a_1\,z + a_2\,z^2 + a_3\,z^3\,,
\end{equation}
without an explicit richness dependence, to simplify the evaluation of the clustering model. The best-fit parameters for the different regions are shown in Table~\ref{table:photoz_params}.

\begin{table}[t]
\centering    
\small
\caption{Best-fit parameters for the different calibrations of the photo-$z$ uncertainty coefficients (Eq.\,\ref{eq:sigma_z}).}
\centering
\begin{tabular}{l c c c c}  
\hline  
Region  &  $a_0$ & $a_1$ & $a_2$ & $a_3$ \\
\hline
North   & 0.00826 & $-$0.01524 & 0.02172 & $-$0.00554 \\
South   & 0.01582 & $-$0.03343 & 0.03688 & $-$0.00936 \\
Mean    & 0.01204 & $-$0.02434 & 0.02930 & $-$0.00745 \\
\hline
\end{tabular}

\label{table:photoz_params} 
\end{table}

\begin{table}[h]
\centering     
\small
\caption{Best-fit values for the different calibrations of the richness-mass relation parameters (Eqs.\,\ref{eq:mor_var} and\,\ref{eq:mor_mean}).}
\centering
\begin{tabular}{l c c c c c}  
\hline  
Region  &  $\logten M_1$ & $\logten M_{\rm min}$ & $\alpha$ & $\epsilon$ & $\sigma_{\rm intr}$ \\
\hline
North   & 11.0889 & 10.0451 & 0.5101 & 1.1003 & 0.4777 \\
South   & 11.1207 & 10.0602 & 0.5095 & 1.3053 & 0.4515 \\
Mean    & 11.1048 & 10.0527 & 0.5099 & 1.2028 & 0.4646 \\
\hline
$\Delta$ & 0.0318 & 0.0151 & $-$0.0005 & 0.2051 & $-$0.0261 \\  
\hline
\end{tabular}
\tablebib{$\Delta$ indicates the difference between the south and north regions for each parameter, used to generate more extreme selection functions. Notation: $\logten M_{\rm x} \equiv \logten [M_{\rm x}/(h^{-1}\,M_\odot)]$ }
\label{table:plob_params} 
\end{table}

In Eq.\,\eqref{eq:pk_a_i}, the expected number density of haloes and the effective bias are given by
\begin{equation}
\label{eq:massfunc_obs}
    n(z, \Delta \lob) = \int_0^\infty \diff M \; \frac{\diff n}{\diff M}(M, z) \int_{\Delta \lob_i} \diff \lob \, {\cal P}(\lob \,|\, M, z)\,,
\end{equation}
\begin{align}
\label{eq:effbias_obs}
    b_{\rm eff}(z, \Delta \lob) = \frac{1}{n(z,\Delta \lob)}\int_0^\infty &\diff M \; \frac{\diff n}{\diff M}(M, z)\,b(M, z)\,\nonumber \\ &\times \int_{\Delta \lob_i} \diff \lob \, {\cal P}(\lob \,|\, M, z)\,,
\end{align}
respectively, where $\diff n/ \diff M$ is the halo mass function, $b$ is the halo bias, and ${\cal P}(\lob \,|\, M, z)$ is the observed richness--mass relation, describing the relation between the observed richness and the true mass and redshift. Such a relation has been calibrated on a \Euclid Flagship 2 mock subsample covering $400\,\deg^2$, split into $200\,\deg^2$ north patch and a $200\,\deg^2$ south patch. The photometric redshift properties of the mock were tuned using the same TR1 reference adopted for the photo-$z$ scatter calibration described above. 
Haloes in the Flagship 2 lightcone are matched to detected cluster catalogues generated independently in the north and south patches with the two cluster finders adopted in the \Euclid cluster-detection pipeline: AMICO \citep{Bellagamba:2017uyp} and PZWav \citep{2014becs.confE...7G}. For each region and finder, the detected-cluster catalogue is joined to the corresponding member catalogue, and member galaxies are associated with their parent Flagship 2 haloes. The cluster--halo association is defined through a percolation-style scheme: for each detection, we consider members with valid membership probability, within the membership radius and satisfying the adopted $H-H^\star$ cut, and assign the cluster to the true halo that contributes the largest summed membership probability. Clusters are processed in order of decreasing observed richness, with each true halo allowed to match at most one detection, avoiding duplicate assignments to lower-richness systems. The conditional distribution ${\cal P}(\lob|M,z)$ is then fitted using the true mass and redshift of the matched halo together with the observed richness $\lob$ measured by the
cluster finder.  The relation follows a continuous-shifted Poisson distribution,
\begin{equation} \label{eq:plob_m}
    P(\lob | M,z) = \frac{{\rm e}^{-\nu} \nu^x}{\Gamma(x+1)}\,,
\end{equation}
Where $\Gamma$ is the Gamma function. The distribution variance, $\nu$, and shifted independent variable, $x$, read as
\begin{equation} \label{eq:mor_var}
    \nu = \langle \lambda^{\rm sat} \rangle + \left[ \sigma_{\rm intr} \langle \lambda^{\rm sat} \rangle \right]^2 \, , \quad  x = \lob - 1 + \left[ \sigma_{\rm intr} \langle \lambda^{\rm sat} \rangle \right]^2 \,,
\end{equation}
where the mean satellite richness is
\begin{equation} \label{eq:mor_mean}
    \langle \lambda^{\rm sat} \rangle =  \left(\frac{M - M_{\rm min}}{M_1 - M_{\rm min}}\right)^\alpha \, \left(\frac{1+z}{1+z_p}\right)^\epsilon \,.
\end{equation}
The parameters $\alpha$, $\epsilon$, $M_1$, $M_{\rm min}$, and $\sigma_{\rm intr}$ are fitted from the matched Flagship 2 cluster--halo catalogues described above. This functional form well approximates the convolution of a Poissonian and Gaussian distribution with zero mean and variance $\delta = \left[ \sigma_\lambda \langle \lambda^{\rm sat} \rangle \right]^2$ \citep[e.g.][]{DES:2018crd}, and correspond to the prescription adopted in \citet{EP-Ingrao}. This approximation is adequate for the purpose of this work, where we are mainly interested in propagating the north--south variations of the observed richness calibration into the clustering model. As noted above, the calibration is performed for AMICO and PZWav, and therefore partially depends on the adopted detection algorithm. However, the calibrated richness distributions obtained with the two finders are sufficiently similar for the analyses presented in this work. We therefore adopt the AMICO calibration as our fiducial choice, and report the corresponding best-fit values in Table~\ref{table:plob_params}.

To take into account the RSDs effect and the photo-$z$ uncertainties, the matter power spectrum of Eq.~\eqref{eq:pk_a_i} must be modified. Our modelling incorporates the linear component of RSDs, namely the Kaiser effect \citep{Kaiser:1987qv}, which produces a scale-independent boost in the clustering amplitude. Nonlinear redshift-space effects such as the finger-of-god suppression are much weaker for cluster samples and can be safely neglected \citep{Veropalumbo:2013cua,Sereno:2014eea}.
Photo-$z$ uncertainties have a distinct impact, damping small-scale correlations, enhancing large-scale ones, and smoothing the BAO feature.
The observed power spectrum in the presence of RSDs and photo-$z$ errors becomes \citep{Sereno:2014eea,Lesci:2022owx}
\begin{equation}
    P_{\rm m}^{\,\rm obs}(k) = (A_1 + \beta\,A_2 + \beta^2\,A_3)\,P_{\rm m}(k)\,,
\end{equation}
where $\beta = f_{\rm g}/b_{\rm eff}$, with $f_{\rm g}$ denoting the linear growth rate, which in a $\Lambda$CDM cosmology is accurately approximated as $f_{\rm g} \simeq \left[\Omega_{\rm m}(z)\right]^{0.55}$, and the coefficients are
\begin{align} 
    A_1 &= \frac{\sqrt{\pi}}{2\,k \sigma}\,{\rm erf}(k \sigma)\,, \label{eq:pk_1}\\ 
    A_2 &= \frac{1}{(k\sigma)^3}\,\bigg[ \frac{\sqrt{\pi}}{2}\,{\rm erf}(k \sigma) - k\sigma\,\exp(-k^2\,\sigma^2)\bigg]\,,  \label{eq:pk_2}\\
    A_3 &= \frac{1}{(k\sigma)^5}\,\bigg[ \frac{3\sqrt{\pi}}{8}\,{\rm erf}(k \sigma) - \frac{k\sigma}{4}\,(2k^2\sigma^2 + 3) + \exp(-k^2\,\sigma^2)\bigg]\,, \label{eq:pk_3}
\end{align}
where $\sigma$ depends on the photo-$z$ uncertainty $\sigma_z$ as
\begin{equation}
    \sigma = \frac{c\,\sigma_z}{H(z)}\,.
\end{equation}

Subdominant distortion effects that requires to be corrected in the power spectrum modelling arise from nonlinear BAO damping and geometric distortions. Such corrections follow the same prescriptions adopted in \citet{Fumagalli25}, to which we refer for a complete description of the theoretical corrections implemented in the 2PCF modelling.

We describe the 2PCF covariance with the semi-analytical model presented in \citet{Fumagalli-EP27}. The covariance in the $a$-th redshift bin and between the $i$-th and $j$-th separation bins is given by
\begin{equation}
    \label{eq:cov_cl}
    \begin{split}
     C_{aij}& =  \frac{1}{V_a} \int_0^\infty \frac{\diff k\,k^2}{2 \pi^2}  \left[ P_{\rm h}(k,\Delta \zob_a) + \frac{1}{n(\Delta \zob_a)} \right ]^2 \,W_i(k)\,W_j(k)\\
    & +  \frac{1}{V_a} \int_0^\infty \frac{\diff k\,k^2}{2 \pi^2}\ P_{\rm h}(k,\Delta \zob_a) \, \left[\frac{1}{n(\Delta \zob_a)}\right]^2\,\frac{W_j(k)}{V_i}\,. 
    \end{split}
\end{equation}
The first integral represents the standard Gaussian covariance, while the second one is the low-order non-Gaussian contribution.
To better reproduce the numerical covariance, the bias and shot-noise terms in the clustering covariance model must be modified by three parameters $\{\alpha_{\rm c},\,\beta_{\rm c},\,\gamma_{\rm c}\}$, such that
\begin{align*}
   b &\longrightarrow \beta_{\rm c} \, b\,,\\
   \frac{1}{n} &\longrightarrow \frac{1+\alpha_{\rm c}}{n} \ \text{in the Gaussian term}\,,\\
   \frac{1}{n} &\longrightarrow \frac{1+\gamma_{\rm c}}{n} \ \text{in the non-Gaussian term}.
\end{align*}
These parameters, whose reference values are $\alpha_{\rm c}=0$, $\beta_{\rm c}=1$, and $\gamma_{\rm c}=0$, can be calibrated from a relatively small number of simulations \citep[of order $10^2$;][]{Fumagalli:2022plg}. Such a calibration can be performed at fixed cosmology, as the three parameters show a negligible cosmology dependence \citep{Fumagalli:2025twg}.

\section{Simulated data} \label{sec:data}
In this section, we describe the simulated lightcones employed in our analysis. We outline the mask geometries used to assess the impact of the survey mask, together with the homogeneous regions adopted to investigate the effects of inhomogeneous sky coverage. We also detail the procedure used to measure the 2PCF and to estimate its covariance matrix.

\subsection{Simulations} \label{sec:data:mocks}

\begin{figure}[t]
    \centering
    \includegraphics[width=0.45\textwidth]{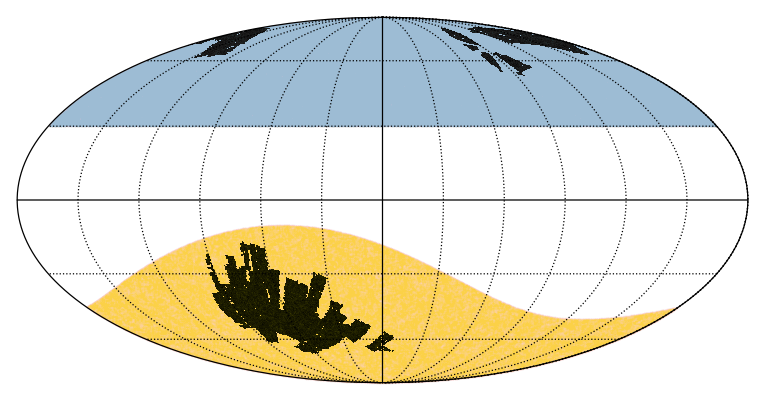}
    \includegraphics[width=0.45\textwidth]{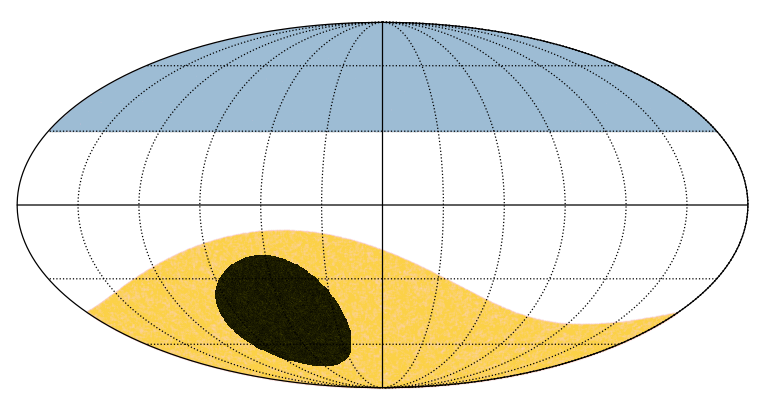}
    \caption{Survey masks. Black areas corresponds to the DR1 mask (\textit{top panel)} and to the corresponding circular mask (\textit{bottom panel}). Coloured regions correspond to the original (blue) and rotated (orange) lightcone footprint.}
    \label{fig:dr1_mask}
\end{figure}

In this work we use 1000 mock catalogues generated with the \texttt{PINOCCHIO} code \citep{Monaco:2001jg,Munari:2016aut}. \texttt{PINOCCHIO} is a fast, semi-analytic algorithm that generates realistic dark matter halo catalogues by combining higher-order Lagrangian perturbation theory with an ellipsoidal-collapse prescription, enabling efficient production of large light-cone simulations. The original lightcones\footnote{The original set of mocks, labelled “NewClusterMocks”, can be found at \url{http://adlibitum.oats.inaf.it/monaco/mocks.html}.} cover $10\,313\,\deg^2$ and redshifts $0 \le z \le 2.5$, and contain haloes with $M_{\rm vir} \ge 10^{13.4}\,h^{-1}\,M_\odot$ in a flat $\Lambda$CDM cosmology from \citet{Planck:2013pxb}: $\Omega_{\rm m } = 0.30711$, $\Omega_{\rm b} = 0.048254$, $h = 0.6777$, $n_{\rm s} = 0.96$, $\sum m_\nu = 0.06\,{\rm eV}$, $A_{\rm s} = 2.21 \times 10^{-9}$, and $\sigma_8 = 0.8288$. Halo masses are rescaled to match the reference halo mass function from \citet{Castro-EP24}. Consistently with the halo mass function model, we use the halo bias model from \citet{EP-Castro24}, calibrated on $N$-body simulations with percentage accuracy. Note that, despite the mass calibration, the bias model is not in perfect agreement with the bias of the simulations \citep{Fumagalli25,EP-Tsaprazi}, due to the approximate nature of the \texttt{PINOCCHIO} code. However, for the level of accuracy of the DR1 data, this residual inconsistency is not relevant.

To mimic realistic observational conditions, each halo is assigned a richness using the mass-richness relation of Eq.\,\eqref{eq:plob_m}, and an observed redshift generated by adding a Gaussian scatter (Eq.\,\ref{eq:photoz}) to the true redshift. The parameters of Eqs.\,\eqref{eq:mor_var},\,\eqref{eq:mor_mean}, and\,\eqref{eq:photoz} are chosen from Tables\,\ref{table:photoz_params} and \ref{table:plob_params}, depending on the  homogeneous regions configuration.

\subsection{Survey mask} \label{sec:data:mask}

To mimic the \Euclid\,DR1 footprint, we applied the corresponding angular mask\footnote{\url{https://www.cosmos.esa.int/web/euclid/euclid-dr1-expected-coverage}} to our lightcones. The expected DR1 coverage spans approximately $500~\deg^2$ in the northern hemisphere and $1600~\deg^2$ in the southern hemisphere, as shown in the top panel of Fig.\,\ref{fig:dr1_mask}. Since our lightcones cover a limited portion of the sky, a single simulation cannot cover the full extent of the mask. We therefore stack together two realizations: one aligned with the north pole (blue region in the figure) and a second one rotated to overlap the southern portion of the mask (orange region in the figure). We obtained 1000 large mocks by stacking one catalogue with the next after rotating the latter; in this way, each mock is used twice, once in the original orientation and once rotated. The survey mask is applied after stacking. The resulting catalogues are independent, since the mask selects different regions of the same lightcone in each case, preventing correlations despite the double use of each individual mock.

As our data set includes only the angular part of the survey mask, we treat the redshift coverage as uniform. To better approximate the observational conditions expected for DR1, we restrict the analysed redshift range to 
$z \in [0.2, 1]$. Within this interval, our masked catalogues contain approximately $10^5$ haloes with virial mass above $10^{13.5}\,h^{-1}\,M_\odot$. The 2PCF measurement uses a subset of these objects, the size of which depends on the richness threshold and varies with the selection function.

As shown by the DR1 case, a realistic survey mask can be characterized by an irregular overall shape, the presence of separate patches, and the occurrence of holes. In principle, the effect of the footprint geometry is automatically accounted for by applying the same mask to the random catalogue. However, irregularities in the mask coverage and the presence of multiple disconnected patches can still impact the measured clustering signal.
To quantify the impact of the survey mask on clustering measurements, we compare the irregular DR1-like mask with a uniform, single-patch circular mask covering the same area (bottom panel of Fig.~\ref{fig:dr1_mask}). This type of mask corresponds to the idealized case typically assumed in theoretical predictions, which do not explicitly model the geometry of the survey mask. The photometric redshift and observed richness are assigned to the haloes within the footprints assuming a homogeneous coverage of the whole sky; the parameters of the photo-$z$ error and richness-mass relation  corresponds to the "Mean" calibration case of Tables\,\ref{table:photoz_params} and\,\ref{table:plob_params}, respectively.

\subsection{Homogeneous regions} \label{sec:data:regions}

\begin{figure}[t]
    \centering
    \includegraphics[width=0.42\textwidth]{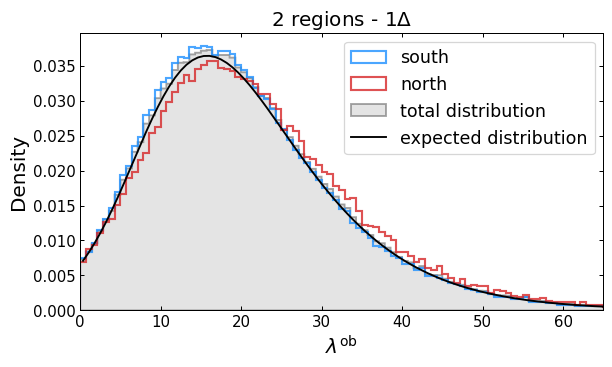}
    \includegraphics[width=0.42\textwidth]{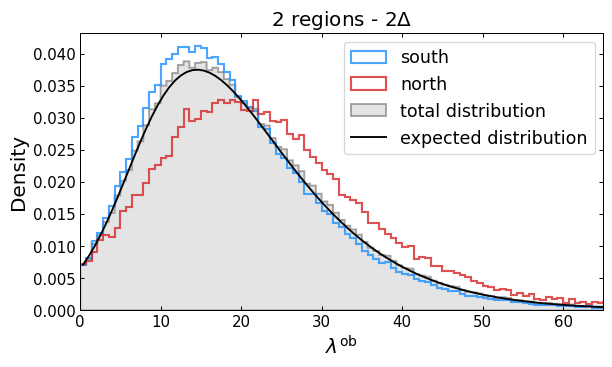}
    \caption{Observed richness distributions for the two homogeneous regions (northern region in red, southern region in blue), compared to the total distribution (grey) and the expected analytical prediction. The \textit{top panel} shows the realistic homogeneous region case, while the \textit{bottom panel} shows the pessimistic case with double shift.}
    \label{fig:dr1_regions}
\end{figure}

In addition to the irregular shape, spatial variations in the survey sensitivity may also influence the measured clustering signal. To explore this, we divide the DR1-like mask into two subregions within which the different observational conditions can be treated as homogeneous. The differences between the homogeneous regions are driven by variations in the photometric redshift quality across the footprint, which affect the inferred cluster richness and can be effectively described through different calibrations of the selection function.

Following the expected DR1 strategy, we consider two regions corresponding to the northern and southern patches of the survey footprint.
Within each region, the observed redshift and richness are assigned to haloes using the corresponding parameter calibrations shown in Tables \ref{table:photoz_params} and\,\ref{table:plob_params}, respectively. The resulting richness distributions are displayed in the upper panel of Fig.~\ref{fig:dr1_regions}. The black curve represents the theoretical prediction, given by the weighted sum of the two distributions; as such, it corresponds to the total distribution that would be obtained if the geometrical subdivision of the mask were ignored, i.e. under the assumption of uniform coverage (grey histogram).

Since a detailed characterisation of the homogeneous regions of the DR1 cluster catalogue is not yet available, the calibration in Table~\ref{table:plob_params} provides a first estimate of the expected level of variation and is used as our reference realistic case. 
To further probe the impact of non-uniform sky coverage, we also investigate more pessimistic scenarios in which the variation between the northern and southern regions is increased. To do so, for each parameter $p_i$ of Eqs.\,\eqref{eq:mor_mean} and \eqref{eq:mor_var}, we compute the difference between the two richness calibrations\footnote{Since the inhomogeneities arise from photo-$z$ calibration, a more accurate approach would be to vary the parameters in Eq.\,\eqref{eq:sigma_z} and propagate the effect to the richness. However, this is computationally demanding, so we directly modify the richness to study the resulting effects.}, 
\begin{equation}
    \Delta = \{p_{{\rm S},i} - p_{\rm N,i} \}\,,
\end{equation}
and generate increasingly extreme configurations by shifting the parameters by a factor $\pm n\Delta_i$, where $n$ is an integer number.
More specifically, in Sect.\,\ref{sec:results:regions} we consider a pessimistic case where the difference between the northern and southern regions is increased by a factor of two ($2\Delta$, bottom panel of Fig.\,\ref{fig:dr1_regions}), which is compared to the realistic scenario ($1\Delta$, corresponding to the default calibration). In Sect.\,\ref{sec:results:regions:robustness} we explore more extreme cases.

\subsection{Measurements}\label{sec:data:measure}
For the 2PCF measurement, we adopt two redshift bins with edges $z \in \{0.2, 0.6, 1.0\}$; the bin width of $\Delta z = 0.4$ is sufficiently large to suppress cross-correlations between adjacent bins \citep{Fumagalli-EP27}. A single richness interval is adopted, applying a lower threshold of $\lambda_{\rm min} = 20$. We measure clustering over the separation range $r = 10$\,--\,$200\,h^{-1}\,$Mpc, which spans the standard linear regime ($r \approx 20$\,--\,$130\,h^{-1}\,$Mpc), extends into mildly non-linear small scales, and includes separations beyond the BAO peak and the zero-crossing scale \citep{Prada:2011uz}. This wide range ensures a more comprehensive characterisation of the mask-induced effects on the clustering signal. The 2PCF is measured using the \citet{Landy:1993yu} estimator. 

The random catalogue is built by drawing objects randomly from each mock and merging the resulting subsets after an additional coordinate shuffling step. This procedure ensures that the random points are properly and uniformly distributed throughout the lightcone volume. The random catalogue has a density 20 times higher than that of a single data mock. The same masks described in the previous sections are applied also to the random catalogue, consistently to the one used in the data (see Appendix\,\ref{app:wrong_rand} for further investigation). Similarly, when different homogeneous regions are introduced in the data catalogue, the same region definition is applied to the random catalogue.

In the following, we use the 2PCF obtained by averaging over all 1000 mocks when comparing clustering signals, rather than relying on a single realisation. This choice mitigates the impact of single mock fluctuations  that may not be fully representative of the ensemble behaviour, and provides a clearer view of any possible systematic effects through the ensemble mean. We nevertheless assign to this mean 2PCF the uncertainty of a single catalogue, corresponding to the statistical precision expected for DR1, which is essential to assess the significance of the observed differences at the survey noise level. The full set of 1000 mocks is also used to estimate numerical covariance matrices.

\section{Results} \label{sec:results}
In this section, we present the results of our analyses, in which we evaluate the impact of the irregular angular mask and non-uniform sky coverage on both the 2PCF monopole and its covariance, by comparing the DR1-like and a circular masks.

\subsection{DR1 vs. circular mask} \label{sec:results:mask}

\begin{figure}[t]
    \centering
    \includegraphics[width=0.49\textwidth]{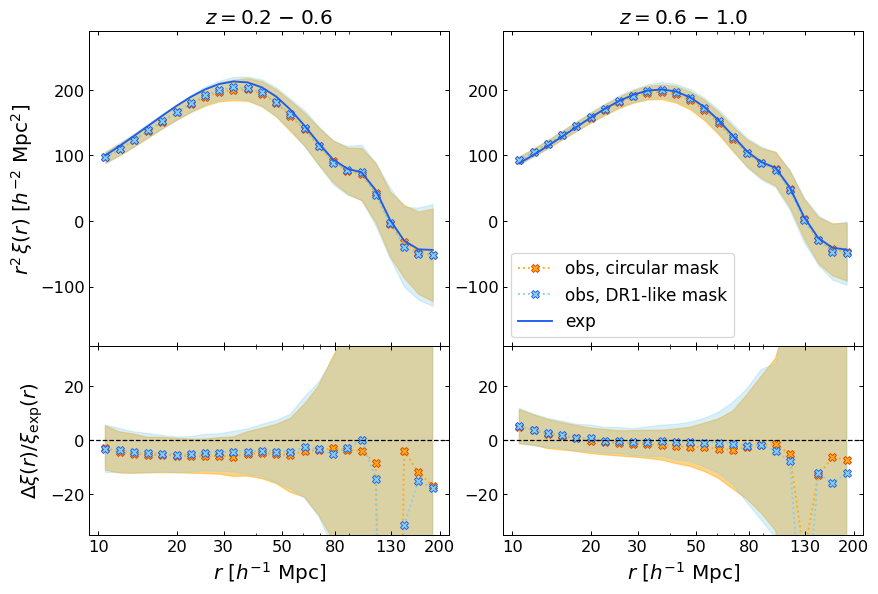}
    \caption{Measured 2PCF of the DR1-like mask (orange) and circular mask (cyan), compared to the expected model (blue line)  of Eq.\,\eqref{eq:2pcf_obs}. Points represent the average measurement over 1000 mocks, shaded areas are the single-mock uncertainty. Percent residuals with respect to the model in the bottom panel. Measurements for the low and high redshift are shown in the \textit{left} and \textit{right} panel, respectively.}
    \label{fig:2pcf_mask}
\end{figure}

\begin{figure}[t]
    \centering
    \includegraphics[width=0.50\textwidth]{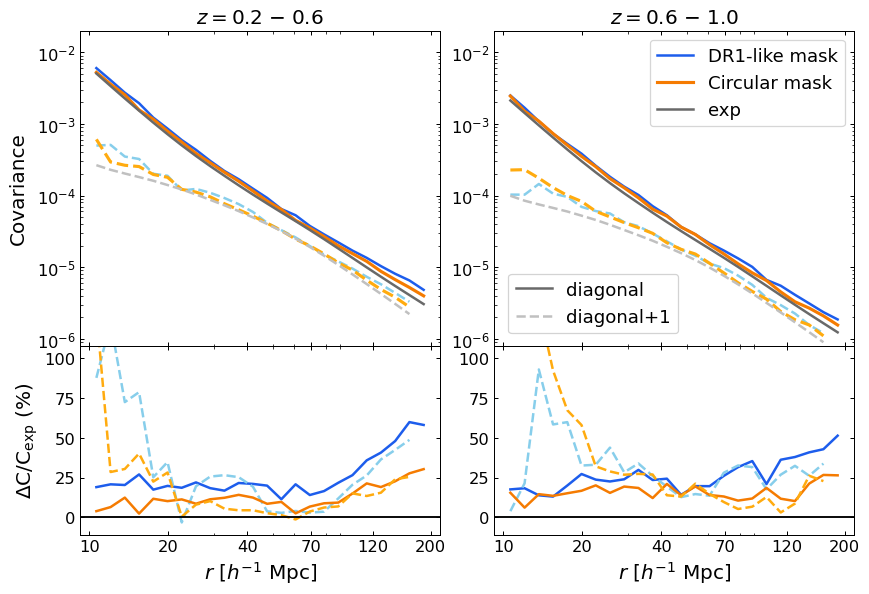}
    \caption{Same as Fig.\,\ref{fig:2pcf_mask} for the 2PCF covariance. Solid lines represent the diagonal terms, while dashed lines are the first off-diagonal elements.}
    \label{fig:cov_mask}
\end{figure}

The first test to evaluate the impact of the survey mask is to determine whether its irregular geometry affect the 2PCF measurement and its covariance, or if properly applying the mask in the random catalogue is sufficient to account for it. To this end, we compare the DR1-like mask with the regular circular mask described in Sect.\,\ref{sec:data:mask}. Figure\,\ref{fig:2pcf_mask} shows the 2PCF monopole measured from the same mock under the two masking schemes. The measurements are in excellent agreement across all scales, well within the DR1-like uncertainty. Consequently, the 2PCF measured with the DR1-like mask is also consistent with the theoretical model of Eq\,\eqref{eq:2pcf_obs}, which is based on the circular-mask case, indicating that no additional mask-related correction is required for modelling the 2PCF shape.

A similar comparison is shown in Fig.~\ref{fig:cov_mask} for the 2PCF covariance. The blue and orange residuals show no significant differences between the two masking cases, except for an almost constant offset in the diagonal element of the DR1-mask case. We can also observe a slightly larger amplitude in the irregular mask case at large scales. This is probably due to border effects that reduce the effective number of objects on the mask edges, especially in the low redshift bin where the sampled volume is smaller. Nevertheless, this deviation is small and affects scales that are usually non relevant for cluster clustering constraints \citep[$r \gtrsim 130\, h^{-1}\,$Mpc, for details, see][]{Fumagalli-EP27}.
In both masking configurations, the numerical covariances show a residual discrepancy with respect to the analytical model, requiring the calibration of the nuisance parameters $\{\alpha_c,\beta_c,\gamma_c\}$ introduced in Sect.~\ref{sec:theory}. These parameters were previously shown to accurately capture the deviations in the regular mask case \citep{Fumagalli-EP27}. Given the close agreement between the covariance matrices of the regular and irregular masking configurations, we expect the same parametrization to remain valid for the irregular mask as well, provided that the calibration is performed applying the actual mask to the synthetic catalogues. Further considerations about the covariance fit are discussed in Sect.\,\ref{sec:results:cov_fit}.

\subsection{Homogeneous regions}\label{sec:results:regions}
The second potential source of mask-related bias in clustering analyses is the non-homogeneity of the sky coverage, i.e., the presence of distinct spatial regions with different observational properties. To test this, we measure the 2PCF and its covariance in our mocks after applying the different cluster selection function to the two homogeneous regions described in Sect.\,\ref{sec:data:regions}. We then measure the 2PCF over the full footprint, treating the catalogue as a single homogeneous sample, without explicitly separating the contributions from the different regions. The only requirement is that the same homogeneous regions used for the data are consistently applied to the random catalogue. This measurement is compared to the 2PCF obtained using a homogeneous mask, with a selection function defined as the average of the two cases.

\begin{figure}[t]
    \centering
    \includegraphics[width=0.50\textwidth]{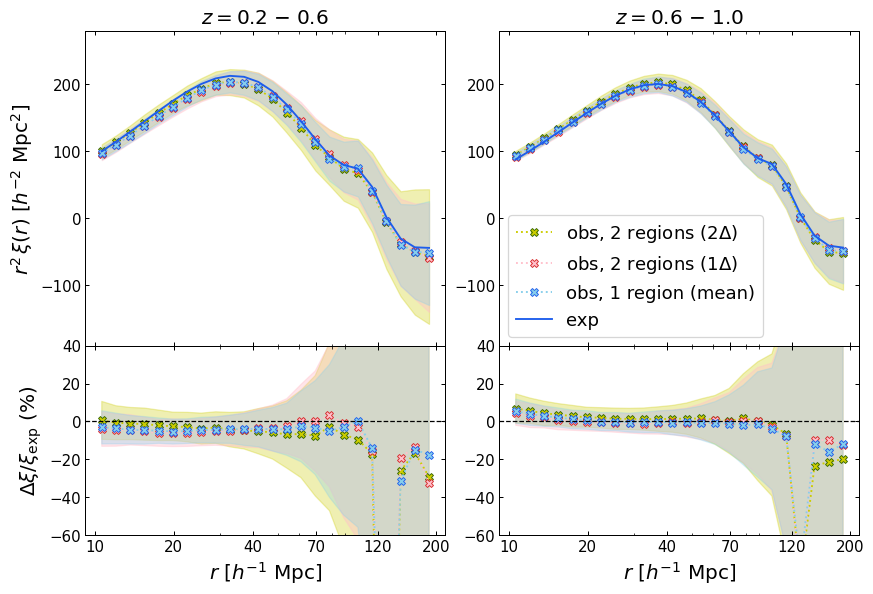}
    \caption{Measured 2PCF for the DR1-like mask divided into two regions, north and south, for the realistic (red) and pessimistic (green) cases, compared to the single region case (cyan). The blue line is the expected model, corresponding to the single region case. Points represent the average measurement over 1000 mocks, shaded areas are the single-mock uncertainty. Percent residuals with respect to the model in the bottom panel. Low-redshift bin in the \textit{left} panel, high-redshift bin in the \textit{right} panel.}
    \label{fig:2pcf_regions}
\end{figure}

\begin{figure}[t]
    \centering
    \includegraphics[width=0.50\textwidth]{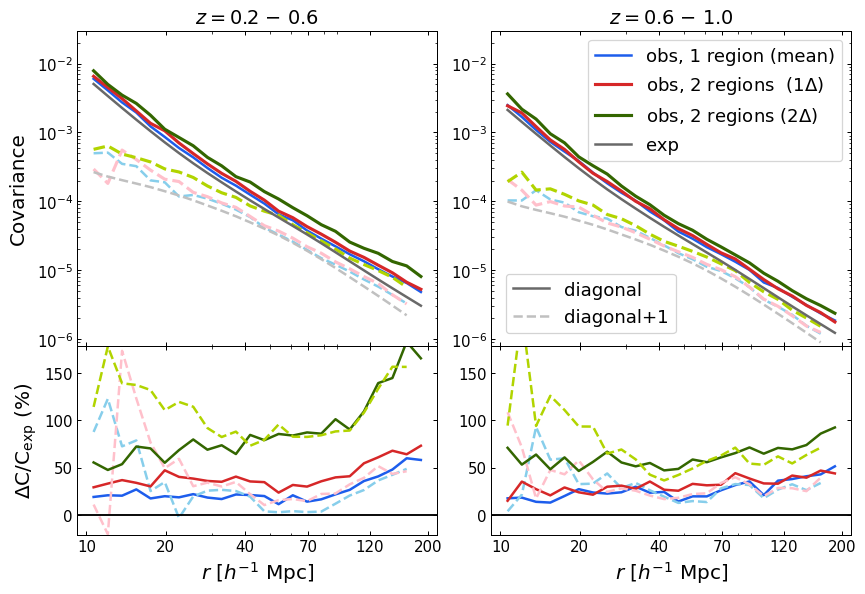}
    \caption{Same as Fig.\,\ref{fig:2pcf_regions} for the 2PCF covariance matrix.  Solid lines represent the diagonal terms, while dashed lines are the first off-diagonal elements.}
    \label{fig:cov_regions}
\end{figure}

Figure~\ref{fig:2pcf_regions} shows that both the realistic and pessimistic cases are consistent with the mean-region measurement, and therefore are in good agreement with the predicted 2PCF, which by construction describes the uniform case. On the contrary, Fig.~\ref{fig:cov_regions} shows clear deviations in the covariance matrix when the sky coverage is spatially inhomogeneous, especially at low redshift where the light cone volume is small and volume effects may become relevant. In particular, the presence of regions with different selection function calibrations introduces an offset with respect to the homogeneous case (green and red lines compared to the blue ones), whose amplitude increases with the discrepancy between the two calibrations. This effect impacts both diagonal and off-diagonal covariance elements, demonstrating that the presence of different regions does not simply modify the number of objects -- and therefore the shot-noise contribution -- but affects the full clustering covariance. When comparing the numerical covariances to the analytical model, we note that the offsets remain approximately scale-independent over the range of scales typically used in clustering analyses, while scale-dependent deviations appear only at the largest separations, probably reflecting the same border effects already observed for the irregular-mask case (Sect.\,\ref{sec:results:mask}). Also in this case, we show that these effects can be effectively absorbed through an appropriate calibration of the covariance model parameters (Sect.\,\ref{sec:results:cov_fit}).

\subsection{Robustness tests on region definitions} \label{sec:results:regions:robustness}
\begin{figure}[t]
    \centering
    \includegraphics[width=0.50\textwidth]{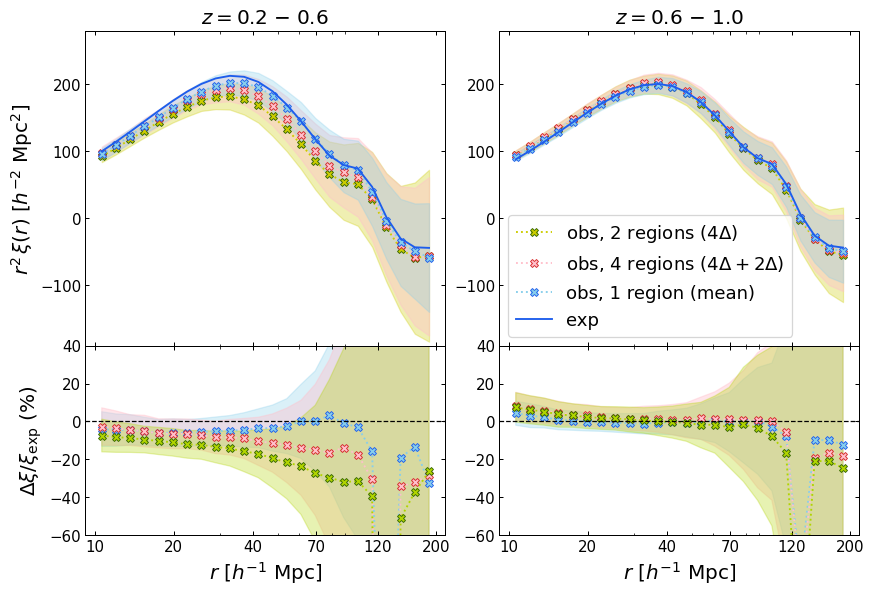}
    \caption{Same as Fig.~\ref{fig:2pcf_regions} for different selection functions and number of regions. Blue points show the uniform mask case, green points the two-region case with calibration shift $4\Delta$, and red points the four-region case with two regions with $4\Delta$ and two regions with $2\Delta$.}
    \label{fig:2pcf_robust}
\end{figure}

To further assess when the presence of different homogeneous regions becomes relevant for clustering analyses, we explore more extreme configurations in which the calibration differences between different regions are progressively increased. In particular, we aim to identify the level of inhomogeneity at which the impact on the 2PCF becomes non-negligible at the DR1 level of accuracy.
We start seeing some significant disagreement when the calibration difference is around $4\Delta$ (green points in Fig.~\ref{fig:2pcf_robust}), producing an underestimation of clustering amplitude which increases with radial separation.
The effect is redshift dependent, and becomes stronger at low redshift. This behaviour may be related to volume effects, since the same level of selection function variation can translate into a larger relative impact when the sampled volume is smaller. To obtain a significant difference in the high redshift bins, we need a variation around $7\Delta$ in the selection functions. 

We also investigate whether, at fixed maximum calibration difference, increasing the number of regions modifies the clustering signal. We therefore create four regions by splitting both the north and south patches in two. We then assign a $4\Delta$ difference to one of the south and north sub-patches and $2\Delta$ difference to the remaining ones. We see from the red points in Fig.\,\ref{fig:2pcf_robust} that the difference is milder than the two region case with $4\Delta$ difference. This means that the effect is primarily driven by the total amplitude of the variation rather than by the number of regions themselves. Indeed, when the same maximum calibration difference is distributed across a larger number of regions -- introducing additional intermediate cases with smaller deviations -- the resulting impact on the 2PCF is reduced. This indicates that what matters is the overall variation of the selection function, rather than the number of homogeneous regions.

\subsection{Covariance fit} \label{sec:results:cov_fit}

\begin{figure}[t]
    \centering
    \includegraphics[width=0.42\textwidth]{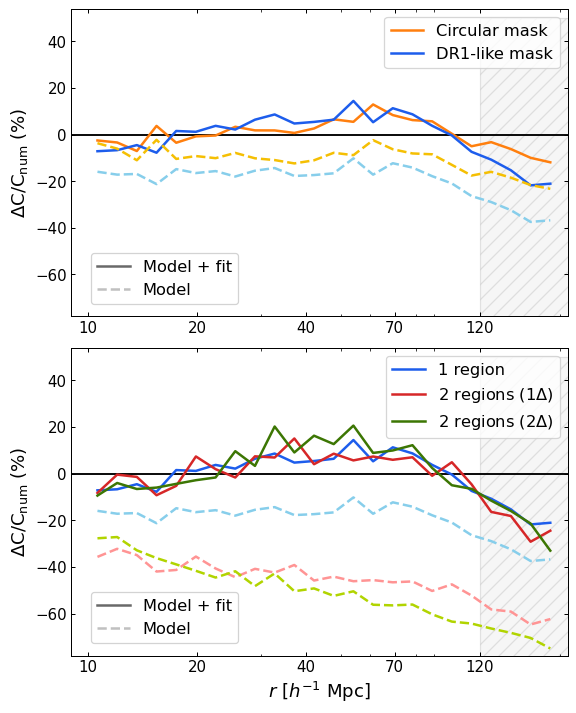}
    \caption{Percent residuals of the diagonal model covariance with respect to the numerical one. Dashed and solid lines represent the semi-analytic model of Eq.\,\eqref{eq:cov_cl}, without and with best-fit parameters, respectively. Shaded grey area denotes the scales excluded from the fit. \textit{Top panel}: DR1-like case (blue) compared to the circular mask case (orange). \textit{Bottom panel}: single region case (blue), compared to the realistic ($1\Delta$, red) and pessimistic ($2\Delta$, green) homogeneous region cases.}
    \label{fig:cov_fit}
\end{figure}

\begin{figure}[t]
    \centering
    \includegraphics[width=0.47\textwidth]{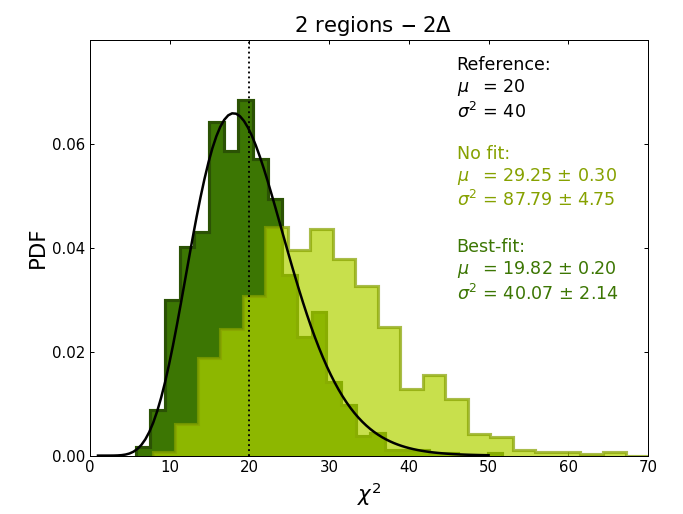}
    \caption{$\chi^2$ distribution of the original (light green) and fitted (dark green) expected covariance, compared to the predicted distribution (black). The errors on the mean $\mu$ and the variance $\sigma^2$ of the distributions are estimated with bootstrap method.}
    \label{fig:chi2_fit}
\end{figure}

In this section, we assess whether the covariance model introduced in Eq.\,\eqref{eq:cov_cl} is still able to describe the numerical covariance with sufficient accuracy in the presence of irregular mask and non-uniform sky coverage. As discussed in the previous sections, the mask geometry introduces a constant offset across over most scales, with departures from this behaviour arising only at large scales. The presence of homogeneous regions primarily increases the amplitude of this offset, while introducing only a mild scale dependence as the variation between the selection functions increases, mainly at low redshift. We fit the semi-analytical model to the numerical covariance measured in each configuration (\citealt{Fumagalli:2022plg} provide details of the fitting procedure), excluding from the fit the largest scales ($r>120\,h^{-1}\,$Mpc), where these scale-dependent deviations start to become significant. Figure~\ref{fig:cov_fit} presents the residuals between the analytical model and the numerical covariance, restricted to the diagonal terms, which provide the dominant contribution. The dashed lines show the prediction obtained using the reference parameters  ($\alpha_c=0$,  $\beta_C=1$, $\gamma_c=0$), while the solid lines show the result after fitting the covariance model to the numerical matrix. The upper panel focuses on the effect of the mask geometry, while the lower panel considers non-uniform sky coverage.
In all cases, the fitted model provides an unbiased description of the numerical covariance, with residual scale-dependent deviations limited to below 10\%.

To assess the quality of the fitted covariance over the full matrix, we compute a $\chi^2$ value for each of the 1000 catalogues, using the catalogue measurements $\boldsymbol{\xi}_i$, the reference prediction $\boldsymbol{\xi}_{\rm ref}$, and the predicted covariance $\mathbf{C}_{\rm model}$:
\begin{equation}
\chi^2_i =
\left(\boldsymbol{\xi}_i-\boldsymbol{\xi}_{\rm ref}\right)^{\rm T}
\mathbf{C}_{\rm model}^{-1}
\left(\boldsymbol{\xi}_i-\boldsymbol{\xi}_{\rm ref}\right)\,.
\end{equation}
As described in \citet{Fumagalli:2022plg}, the  $\chi^2$ distribution resulting from a well-estimated covariance matrix is expected to be consistent with the reference distribution. We show the results for the most extreme case considered above, namely two homogeneous regions with $\Delta=2$. While the original analytical covariance matrix produces a $\chi^2$ distribution that significantly deviates from the expectation, the best-fit covariance brings it into excellent agreement. This demonstrates that the fit provides an accurate description not only of the diagonal terms, but of the full covariance matrix.

\section{Conclusions} \label{sec:conclusions}
In this work, we assessed how realistic survey masks affect cluster clustering measurements in a DR1-like \Euclid survey. Using a suite of lightcone mocks generated with the \texttt{PINOCCHIO} algorithm, we tested both the impact of the mask geometry and the effect of spatially varying selection functions. Our goal was to determine whether such effects can bias the 2PCF and its covariance, and to identify the level of modelling required for robust cosmological analyses.
Our main findings are as follows:
\begin{itemize}
    \item The complex geometry of the mask does not bias the 2PCF monopole when the same footprint is applied to the random catalogue. The measured 2PCF remains fully consistent with the theoretical model without the need of further corrections;
    \item Spatial inhomogeneities across the footprint do not bias the 2PCF amplitude at the level of variation expected for the DR1 selection functions. To obtain appreciable effects on the 2PCF, the amount of discrepancy between the selection functions in different regions must be increased by a significant factor compared to the current expectation: around 4 at low redshift and more than 7 at high redshift;
    \item When appreciable, the impact of spatial inhomogeneity is driven by the degree of variation among the selection functions rather than by the number of regions;
    \item The irregular mask introduces a small increase in the diagonal elements of the 2PCF covariance. Inhomogeneous selection functions increase not only the shot noise but also the overall amplitude of the covariance, including off-diagonal terms. In both cases, the induced change is nearly scale-independent and can be corrected through the calibration of the covariance model of Eq.\,\eqref{eq:cov_cl} with the correct mask and sky coverage applied. 
\end{itemize}

Our analysis shows that the geometry and inhomogeneous structure of the survey mask will not require any extra modelling for the analysis of DR1 cluster data.
However, it is important to notice that such effect could introduce biases in cluster clustering for future surveys, as well as for future \Euclid data releases, which will be characterized by larger samples and therefore higher accuracy. This calls for dedicated corrections in the modelling that are conceptually simple, yet overlooking or mismodelling them risks introducing systematic errors in the inferred cosmological parameters. To ensure accurate and unbiased cosmological inferences, mask-induced effects must be systematically tested and calibrated using realistic mocks tailored to each survey, since the effect cannot be predicted a priori. This will be particularly important as cluster cosmology enters the high-precision era with the Euclid Data Release 2 and, later, with the full \Euclid survey, where even small uncorrected systematics could compromise the gains expected from the unprecedented statistical power of upcoming cluster catalogues.

\begin{acknowledgements}
\AckEC
This work has made use of CosmoHub \citep{Tallada:2020qmg,Carretero:2017zkw}, developed by PIC (maintained by IFAE and CIEMAT) in collaboration with ICE-CSIC. It received funding from the Spanish government (grant EQC2021-007479-P funded by MCIN/AEI/10.13039/501100011033), the EU NextGeneration/PRTR (PRTR-C17.I1), and the Generalitat de Catalunya
\end{acknowledgements}

\bibliographystyle{aa} 
\bibliography{biblio,Euclid}

\begin{thebibliography}{43}
\expandafter\ifx\csname natexlab\endcsname\relax\def\natexlab#1{#1}\fi

\bibitem[{Beauchamps {et~al.}(2022)Beauchamps, Lacasa, Tutusaus, Aubert, Baratta, Gorce, \& Sakr}]{Beauchamps:2021fhb}
Beauchamps, S.~G., Lacasa, F., Tutusaus, I., {et~al.} 2022, \aap, 659, A128

\bibitem[{Bellagamba {et~al.}(2018)Bellagamba, Roncarelli, Maturi, \& Moscardini}]{Bellagamba:2017uyp}
Bellagamba, F., Roncarelli, M., Maturi, M., \& Moscardini, L. 2018, \mnras, 473, 5221

\bibitem[{{Bocquet} {et~al.}(2024){Bocquet}, {Grandis}, {Bleem}, {et~al.}}]{SPT:2024qbr}
{Bocquet}, S., {Grandis}, S., {Bleem}, L.~E., {et~al.} 2024, Phys. Rev. D, 110, 083510

\bibitem[{Borgani {et~al.}(1999)Borgani, Plionis, \& Kolokotronis}]{Borgani:1998sfa}
Borgani, S., Plionis, M., \& Kolokotronis, E. 1999, \mnras, 305, 866

\bibitem[{Borgani {et~al.}(2001)Borgani, Rosati, Tozzi, Stanford, Eisenhardt, Lidman, Holden, Della~Ceca, Norman, \& Squires}]{Borgani:2001ir}
Borgani, S., Rosati, P., Tozzi, P., {et~al.} 2001, \apj, 561, 13

\bibitem[{Carretero {et~al.}(2017)}]{Carretero:2017zkw}
Carretero, J. {et~al.} 2017, PoS, EPS-HEP2017, 488

\bibitem[{Colavincenzo {et~al.}(2017)Colavincenzo, Monaco, Sefusatti, \& Borgani}]{Colavincenzo:2017yhx}
Colavincenzo, M., Monaco, P., Sefusatti, E., \& Borgani, S. 2017, JCAP, 03, 052

\bibitem[{{Costanzi} {et~al.}(2019){Costanzi}, {Rozo}, {Simet}, {et~al.}}]{DES:2018crd}
{Costanzi}, M., {Rozo}, E., {Simet}, M., {et~al.} 2019, \mnras, 488, 4779

\bibitem[{{Costanzi} {et~al.}(2021){Costanzi}, {Saro}, {Bocquet}, {et~al.}}]{DES:2020cbm}
{Costanzi}, M., {Saro}, A., {Bocquet}, S., {et~al.} 2021, \prd, 103, 043522

\bibitem[{{Euclid Collaboration: Castander} {et~al.}(2025){Euclid Collaboration: Castander}, {Fosalba}, {Stadel}, {et~al.}}]{EuclidSkyFlagship}
{Euclid Collaboration: Castander}, F., {Fosalba}, P., {Stadel}, J., {et~al.} 2025, A\&A, 697, A5

\bibitem[{{Euclid Collaboration: Castro} {et~al.}(2023){Euclid Collaboration: Castro}, {Fumagalli}, {Angulo}, {et~al.}}]{Castro-EP24}
{Euclid Collaboration: Castro}, T., {Fumagalli}, A., {Angulo}, R.~E., {et~al.} 2023, \aap, 671, A100

\bibitem[{{Euclid Collaboration: Castro} {et~al.}(2024){Euclid Collaboration: Castro}, {Fumagalli}, {Angulo}, {et~al.}}]{EP-Castro24}
{Euclid Collaboration: Castro}, T., {Fumagalli}, A., {Angulo}, R.~E., {et~al.} 2024, \aap, 691, A62

\bibitem[{Euclid Collaboration:~Fumagalli {et~al.}(2021)Euclid Collaboration:~Fumagalli, Saro, Borgani, {et~al.}}]{Euclid:2021api}
Euclid Collaboration:~Fumagalli, A., Saro, A., Borgani, S., {et~al.} 2021, \aap, 652, A21

\bibitem[{{Euclid Collaboration: Fumagalli} {et~al.}(2024){Euclid Collaboration: Fumagalli}, {Saro}, {Borgani}, {et~al.}}]{Fumagalli-EP27}
{Euclid Collaboration: Fumagalli}, A., {Saro}, A., {Borgani}, S., {et~al.} 2024, \aap, 683, A253

\bibitem[{{Euclid Collaboration: Humphrey} {et~al.}(2025){Euclid Collaboration: Humphrey}, {Cunha}, {Bisigello}, {et~al.}}]{EP-Humphrey}
{Euclid Collaboration: Humphrey}, A., {Cunha}, P.~A.~C., {Bisigello}, L., {et~al.} 2025, \aap, 702, A74

\bibitem[{{Euclid Collaboration: Ingrao} {et~al.}(2026){Euclid Collaboration: Ingrao}, {Borgani}, {Costanzi}, {et~al.}}]{EP-Ingrao}
{Euclid Collaboration: Ingrao}, R., {Borgani}, S., {Costanzi}, M., {et~al.} 2026, A\&A, in press, \url{https://doi.org/10.1051/0004-6361/202660776}, arXiv:2605.02723

\bibitem[{{Euclid Collaboration: Mellier} {et~al.}(2025){Euclid Collaboration: Mellier}, {Abdurro'uf}, {Acevedo~Barroso}, {et~al.}}]{EuclidSkyOverview}
{Euclid Collaboration: Mellier}, Y., {Abdurro'uf}, {Acevedo~Barroso}, J., {et~al.} 2025, A\&A, 697, A1

\bibitem[{{Euclid Collaboration: Scaramella} {et~al.}(2022){Euclid Collaboration: Scaramella}, {Amiaux}, {Mellier}, {et~al.}}]{Scaramella-EP1}
{Euclid Collaboration: Scaramella}, R., {Amiaux}, J., {Mellier}, Y., {et~al.} 2022, \aap, 662, A112

\bibitem[{{Euclid Collaboration: Tsaprazi} {et~al.}(2026){Euclid Collaboration: Tsaprazi}, {Fumagalli}, {Marulli}, {et~al.}}]{EP-Tsaprazi}
{Euclid Collaboration: Tsaprazi}, E., {Fumagalli}, A., {Marulli}, F., {et~al.} 2026, A\&A, in press, \url{https://doi.org/10.1051/0004-6361/202660667}, arXiv:2604.25762

\bibitem[{Fumagalli {et~al.}(2022)Fumagalli, Biagetti, Saro, Sefusatti, Slosar, Monaco, \& Veropalumbo}]{Fumagalli:2022plg}
Fumagalli, A., Biagetti, M., Saro, A., {et~al.} 2022, JCAP, 12, 022

\bibitem[{Fumagalli {et~al.}(2025)Fumagalli, Castro, Borgani, \& Valentini}]{Fumagalli:2025twg}
Fumagalli, A., Castro, T., Borgani, S., \& Valentini, M. 2025, \aap, 697, A140

\bibitem[{{Fumagalli} {et~al.}(2026){Fumagalli}, {Costanzi}, {Castro}, {et~al.}}]{Fumagalli25}
{Fumagalli}, A., {Costanzi}, M., {Castro}, T., {et~al.} 2026, \aap, 709, A102

\bibitem[{Fumagalli {et~al.}(2024)Fumagalli, Costanzi, Saro, Castro, \& Borgani}]{Fumagalli:2023yym}
Fumagalli, A., Costanzi, M., Saro, A., Castro, T., \& Borgani, S. 2024, \aap, 682, A148

\bibitem[{{Ghirardini} {et~al.}(2024){Ghirardini}, {Bulbul}, {Artis}, {et~al.}}]{Ghirardini:2024yni}
{Ghirardini}, V., {Bulbul}, E., {Artis}, E., {et~al.} 2024, \aap, 689, A298

\bibitem[{{Gonzalez}(2014)}]{2014becs.confE...7G}
{Gonzalez}, A. 2014, in Building the Euclid Cluster Survey - Scientific Program, 7

\bibitem[{Kaiser(1987)}]{Kaiser:1987qv}
Kaiser, N. 1987, \mnras, 227, 1

\bibitem[{Kravtsov \& Borgani(2012)}]{Kravtsov:2012zs}
Kravtsov, A. \& Borgani, S. 2012, AR\aap, 50, 353

\bibitem[{Landy \& Szalay(1993)}]{Landy:1993yu}
Landy, S.~D. \& Szalay, A.~S. 1993, \apj, 412, 64

\bibitem[{{Laureijs} {et~al.}(2011){Laureijs}, {Amiaux}, {Arduini}, {Augu{\`e}res}, {Brinchmann}, {Cole}, {Cropper}, {Dabin}, {Duvet}, {Ealet}, {Garilli}, {Gondoin}, {Guzzo}, {Hoar}, {Hoekstra}, {Holmes}, {Kitching}, {Maciaszek}, {Mellier}, {Pasian}, {Percival}, {Rhodes}, {Saavedra Criado}, {Sauvage}, {Scaramella}, {Valenziano}, {Warren}, {Bender}, {Castander}, {Cimatti}, {Le F{\`e}vre}, {Kurki-Suonio}, {Levi}, {Lilje}, {Meylan}, {Nichol}, {Pedersen}, {Popa}, {Rebolo Lopez}, {Rix}, {Rottgering}, {Zeilinger}, {Grupp}, {Hudelot}, {Massey}, {Meneghetti}, {Miller}, {Paltani}, {Paulin-Henriksson}, {Pires}, {Saxton}, {Schrabback}, {Seidel}, {Walsh}, {Aghanim}, {Amendola}, {Bartlett}, {Baccigalupi}, {Beaulieu}, {Benabed}, {Cuby}, {Elbaz}, {Fosalba}, {Gavazzi}, {Helmi}, {Hook}, {Irwin}, {Kneib}, {Kunz}, {Mannucci}, {Moscardini}, {Tao}, {Teyssier}, {Weller}, {Zamorani}, {Zapatero Osorio}, {Boulade}, {Foumond}, {Di Giorgio}, {Guttridge}, {James}, {Kemp}, {Martignac}, {Spencer}, {Walton}, {Bl{\"u}mchen}, {Bonoli},
  {Bortoletto}, {Cerna}, {Corcione}, {Fabron}, {Jahnke}, {Ligori}, {Madrid}, {Martin}, {Morgante}, {Pamplona}, {Prieto}, {Riva}, {Toledo}, {Trifoglio}, {Zerbi}, {Abdalla}, {Douspis}, {Grenet}, {Borgani}, {Bouwens}, {Courbin}, {Delouis}, {Dubath}, {Fontana}, {Frailis}, {Grazian}, {Koppenh{\"o}fer}, {Mansutti}, {Melchior}, {Mignoli}, {Mohr}, {Neissner}, {Noddle}, {Poncet}, {Scodeggio}, {Serrano}, {Shane}, {Starck}, {Surace}, {Taylor}, {Verdoes-Kleijn}, {Vuerli}, {Williams}, {Zacchei}, {Altieri}, {Escudero Sanz}, {Kohley}, {Oosterbroek}, {Astier}, {Bacon}, {Bardelli}, {Baugh}, {Bellagamba}, {Benoist}, {Bianchi}, {Biviano}, {Branchini}, {Carbone}, {Cardone}, {Clements}, {Colombi}, {Conselice}, {Cresci}, {Deacon}, {Dunlop}, {Fedeli}, {Fontanot}, {Franzetti}, {Giocoli}, {Garcia-Bellido}, {Gow}, {Heavens}, {Hewett}, {Heymans}, {Holland}, {Huang}, {Ilbert}, {Joachimi}, {Jennins}, {Kerins}, {Kiessling}, {Kirk}, {Kotak}, {Krause}, {Lahav}, {van Leeuwen}, {Lesgourgues}, {Lombardi}, {Magliocchetti}, {Maguire},
  {Majerotto}, {Maoli}, {Marulli}, {Maurogordato}, {McCracken}, {McLure}, {Melchiorri}, {Merson}, {Moresco}, {Nonino}, {Norberg}, {Peacock}, {Pello}, {Penny}, {Pettorino}, {Di Porto}, {Pozzetti}, {Quercellini}, {Radovich}, {Rassat}, {Roche}, {Ronayette}, {Rossetti}, {Sartoris}, {Schneider}, {Semboloni}, {Serjeant}, {Simpson}, {Skordis}, {Smadja}, {Smartt}, {Spano}, {Spiro}, {Sullivan}, {Tilquin}, {Trotta}, {Verde}, {Wang}, {Williger}, {Zhao}, {Zoubian}, \& {Zucca}}]{Laureijs11}
{Laureijs}, R., {Amiaux}, J., {Arduini}, S., {et~al.} 2011, ESA/SRE(2011)12, arXiv:1110.3193

\bibitem[{Lesci {et~al.}(2025)Lesci, Marulli, Moscardini, {et~al.}}]{Lesci:2025bqt}
Lesci, G.~F., Marulli, F., Moscardini, L., {et~al.} 2025, \aap, 703, A25

\bibitem[{Lesci {et~al.}(2022)Lesci, Nanni, Marulli, {et~al.}}]{Lesci:2022owx}
Lesci, G.~F., Nanni, L., Marulli, F., {et~al.} 2022, \aap, 665, A100

\bibitem[{Mana {et~al.}(2013)Mana, Giannantonio, Weller, Hoyle, Huetsi, \& Sartoris}]{Mana:2013qba}
Mana, A., Giannantonio, T., Weller, J., {et~al.} 2013, \mnras, 434, 684

\bibitem[{Monaco {et~al.}(2002)Monaco, Theuns, \& Taffoni}]{Monaco:2001jg}
Monaco, P., Theuns, T., \& Taffoni, G. 2002, \mnras, 331, 587

\bibitem[{Munari {et~al.}(2017)Munari, Monaco, Sefusatti, Castorina, Mohammad, Anselmi, \& Borgani}]{Munari:2016aut}
Munari, E., Monaco, P., Sefusatti, E., {et~al.} 2017, \mnras, 465, 4658

\bibitem[{Percival {et~al.}(2010)Percival, Reid, Eisenstein, {et~al.}}]{SDSS:2009ocz}
Percival, W.~J., Reid, B.~A., Eisenstein, D.~J., {et~al.} 2010, \mnras, 401, 2148

\bibitem[{{Planck Collaboration XVI.}(2014)}]{Planck:2013pxb}
{Planck Collaboration XVI.} 2014, \aap, 571, A16

\bibitem[{Prada {et~al.}(2011)Prada, Klypin, Yepes, Nuza, \& Gottloeber}]{Prada:2011uz}
Prada, F., Klypin, A., Yepes, G., Nuza, S.~E., \& Gottloeber, S. 2011 [\eprint[arXiv]{1111.2889}]

\bibitem[{Romanello {et~al.}(2025)Romanello, Marulli, Moscardini, Giocoli, Lesci, Contarini, Fumagalli, \& Sartoris}]{Romanello:2024kbo}
Romanello, M., Marulli, F., Moscardini, L., {et~al.} 2025, \aap, 693, A195

\bibitem[{Sato {et~al.}(2013)Sato, H{\"u}tsi, Nakamura, \& Yamamoto}]{Sato:2013hea}
Sato, T., H{\"u}tsi, G., Nakamura, G., \& Yamamoto, K. 2013, Int. J. \aap, 3, 243

\bibitem[{Sereno {et~al.}(2015)Sereno, Veropalumbo, Marulli, Covone, Moscardini, \& Cimatti}]{Sereno:2014eea}
Sereno, M., Veropalumbo, A., Marulli, F., {et~al.} 2015, \mnras, 449, 4147

\bibitem[{Tallada {et~al.}(2020)}]{Tallada:2020qmg}
Tallada, P. {et~al.} 2020, Astron. Comput., 32, 100391

\bibitem[{To {et~al.}(2021)}]{DES:2020uce}
To, C.-H. {et~al.} 2021, \mnras, 502, 4093

\bibitem[{Veropalumbo {et~al.}(2014)Veropalumbo, Marulli, Moscardini, Moresco, \& Cimatti}]{Veropalumbo:2013cua}
Veropalumbo, A., Marulli, F., Moscardini, L., Moresco, M., \& Cimatti, A. 2014, \mnras, 442, 3275

\end{thebibliography}

\begin{appendix}

\section{Wrong random mask} \label{app:wrong_rand}
\begin{figure}[h]
    \centering
    \includegraphics[width=0.5\textwidth]{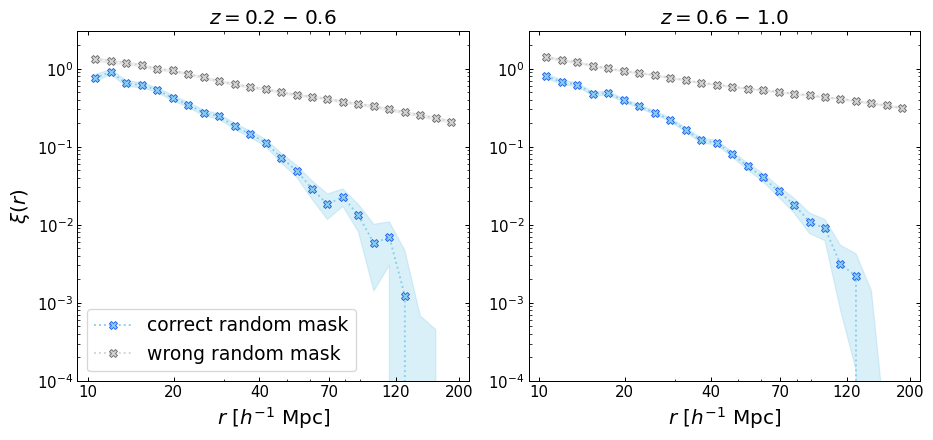}
    \caption{Measured 2PCF for the DR1-like mask case. In the blue case, the DR1-like mask has been applied both to the data and random catalogues, while the grey points use the DR1-like mask for the data and the circular mask for the random.}
    \label{fig:wrong_rand}
\end{figure}

We explicitly show here the effect of using an inconsistent mask for the random catalogue. Figure\,\ref{fig:wrong_rand} shows that applying the DR1-like mask to the data, but the circular mask to the random catalogue significantly distorts the shape of the measured 2PCF. This underscores the importance of matching the geometry of the random catalogue to that of the data. Using a consistent mask in the randoms is an essential step to mitigate geometric effects induced by the survey footprint. Under this assumption, further investigations of the mask related impacts are presented in Sect.\,\ref{sec:results:mask}.

\section{Number counts covariance} \label{app:count_cov}
The covariance of cluster number counts is directly related to clustering through its sample variance term, corresponding to the projection of the clustering signal along the line of sight within the redshift bin \citep[section 2.3 of][]{Euclid:2021api}. The analytical model describing the sample variance contribution assumes azimuthal symmetry, corresponding to a circular survey aperture \citep[Appendix A of][]{DES:2018crd}. Here, we investigate whether an irregular survey mask modifies the measured number count covariance, which would require a more sophisticated modelling of the analytical sample covariance.

\begin{figure}[h]
    \centering
    \includegraphics[width=0.5\textwidth]{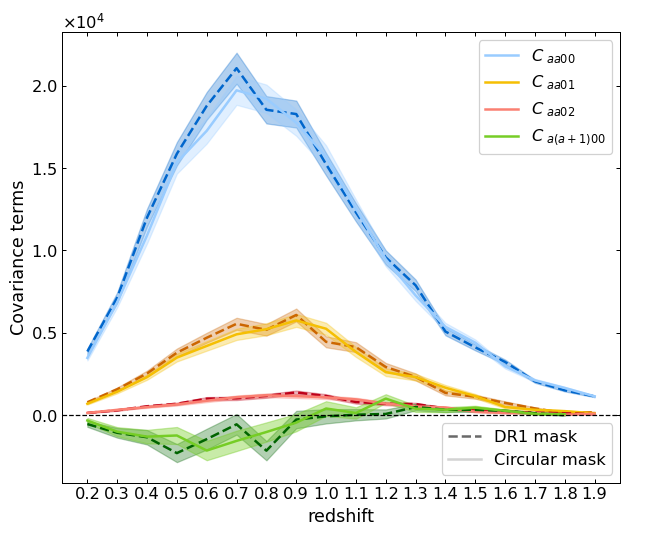}
    \caption{Number count covariance as a function of redshift. Numerical matrix obtained applying the DR1-like (dashed lines) and the circular (solid lines) masks. Different colours represent different components of the covariance: blue is for diagonal elements; orange and red are for the first and second off-diagonal terms between mass bins, respectively; and green is for the first off-diagonal elements in redshift bins.}
    \label{fig:counts_cov}
\end{figure}

While we do not find any appreciable impact of the irregular mask on the 2PCF monopole itself (Sect.\,\ref{sec:results:mask}), its effect on the number counts covariance could, in principle, be different. In particular, as shown in \citet{Fumagalli25}, the presence of photo-$z$ uncertainties induces a dependence of the sample variance contribution on the full anisotropic clustering, including higher-order multipoles beyond the monopole.
We therefore evaluate the counts covariance using the same suite of simulations employed in this work, applying both the irregular and reference mask configurations shown in Fig.\,\ref{fig:dr1_mask}. As shown in Fig.\,\ref{fig:counts_cov}, we find no significant difference between the two cases, indicating that the standard assumption of circular aperture remains valid for the level of accuracy required in this analysis.

\end{appendix}
\end{document}